\documentstyle[epsfig]{mn}

\newif\ifAMStwofonts

\ifoldfss
  \ifCUPmtlplainloaded \else
    \NewTextAlphabet{textbfit} {cmbxti10} {}
    \NewTextAlphabet{textbfss} {cmssbx10} {}
    \NewMathAlphabet{mathbfit} {cmbxti10} {} 
    \NewMathAlphabet{mathbfss} {cmssbx10} {} 
  \fi
  \ifAMStwofonts
    \ifCUPmtlplainloaded \else
      \NewSymbolFont{upmath} {eurm10}
      \NewSymbolFont{AMSa} {msam10}
      \NewMathSymbol{\upi}     {0}{upmath}{19}
      \NewMathSymbol{\umu}     {0}{upmath}{16}
      \NewMathSymbol{\upartial}{0}{upmath}{40}
      \NewMathSymbol{\leqslant}{3}{AMSa}{36}
      \NewMathSymbol{\geqslant}{3}{AMSa}{3E}

      \let\leq=\leqslant 
       
    \fi
  \fi
\fi 

\ifnfssone
  \newmathalphabet{\mathit}
  \addtoversion{normal}{\mathit}{cmr}{m}{it}
  \addtoversion{bold}{\mathit}{cmr}{bx}{it}
  \newmathalphabet{\mathbfit} 
  \addtoversion{normal}{\mathbfit}{cmr}{bx}{it}
  \addtoversion{bold}{\mathbfit}{cmr}{bx}{it}
  \newmathalphabet{\mathbfss} 
  \addtoversion{normal}{\mathbfss}{cmss}{bx}{n}
  \addtoversion{bold}{\mathbfss}{cmss}{bx}{n}
  \ifAMStwofonts
    \ifCUPmtlplainloaded \else
      \UseAMStwoboldmath
      \makeatletter
      \new@mathgroup\upmath@group
      \define@mathgroup\mv@normal\upmath@group{eur}{m}{n}
      \define@mathgroup\mv@bold\upmath@group{eur}{b}{n}
      \edef\UPM{\hexnumber\upmath@group}
      \new@mathgroup\amsa@group
      \define@mathgroup\mv@normal\amsa@group{msa}{m}{n}
      \define@mathgroup\mv@bold\amsa@group{msa}{m}{n}
      \edef\AMSa{\hexnumber\amsa@group}
      \makeatother
      \mathchardef\upi="0\UPM19
      \mathchardef\umu="0\UPM16
      \mathchardef\upartial="0\UPM40
      \mathchardef\leqslant="3\AMSa36
      \mathchardef\geqslant="3\AMSa3E

      \let\leq=\leqslant 

    \fi
  \fi
\fi 

\ifnfsstwo
  \DeclareMathAlphabet{\mathbfit}{OT1}{cmr}{bx}{it}
  \SetMathAlphabet\mathbfit{bold}{OT1}{cmr}{bx}{it}
  \DeclareMathAlphabet{\mathbfss}{OT1}{cmss}{bx}{n}
  \SetMathAlphabet\mathbfss{bold}{OT1}{cmss}{bx}{n}
  \ifAMStwofonts
    \ifCUPmtlplainloaded \else
      \DeclareSymbolFont{UPM}{U}{eur}{m}{n}
      \SetSymbolFont{UPM}{bold}{U}{eur}{b}{n}
      \DeclareSymbolFont{AMSa}{U}{msa}{m}{n}
      \DeclareMathSymbol{\upi}{0}{UPM}{"19}
      \DeclareMathSymbol{\umu}{0}{UPM}{"16}
      \DeclareMathSymbol{\upartial}{0}{UPM}{"40}
      \DeclareMathSymbol{\leqslant}{3}{AMSa}{"36}
      \DeclareMathSymbol{\geqslant}{3}{AMSa}{"3E}

      \let\leq=\leqslant 

    \fi
  \fi
\fi 

\ifCUPmtlplainloaded \else
  \ifAMStwofonts \else 
    \def\upi{\pi}
    \def\umu{\mu}
    \def\upartial{\partial}
  \fi
\fi

\title[PS detection using the SMHW]{Point Source Detection using the Spherical
Mexican Hat Wavelet on simulated all-sky Planck maps}

\author[Vielva et al.]
       {P.~Vielva$^{1,2,}$\footnotemark,
	E.~Mart{\'\i}nez-Gonz{\'a}lez$^{1}$,
        J.~E.~Gallegos$^{1}$,
        L.~~Toffolatti$^{3}$ and \and
	J.~L.~Sanz$^{1}$ \\
	$^{1}$Instituto de F{\'\i}sica de Cantabria, Fac. Ciencias,
	Avda. Los Castros s/n, 39005, Spain \\
        $^{2}$Departamento de F{\'\i}sica Moderna, Universidad de Cantabria,
        Avda. Los Castros s/n, 39005 Santander, Spain\\
	$^{3}$Departamento de F{\'\i}sica, Universidad de Oviedo,
        c/ Calvo Sotelo s/n, 33007 Oviedo, Spain}

\date{\today}

\begin{document}

\maketitle

\label{firstpage}

\begin{abstract}
We present an estimation of the point source (PS) catalogue that could be
extracted from the forthcoming ESA Planck mission data.
We have applied the Spherical Mexican Hat Wavelet (SMHW) 
to simulated all-sky maps that include CMB,
Galactic emission (thermal dust, free-free and synchrotron),
thermal Sunyaev-Zel'dovich effect and PS emission, as well as
instrumental white noise . This work is
an extension of the one presented in Vielva et al. (2001a). We have
developed an algorithm focused on a fast local optimal scale
determination, that is crucial to achieve a PS catalogue with
a large number of detections and a low flux limit. An important effort has
been also done to reduce the CPU time processor for spherical harmonic
transformation, in order to perform the PS detection in a reasonable time.
The presented algorithm is able to provide a PS catalogue above
fluxes: 0.48 Jy (857 GHz), 0.49 Jy (545 GHz), 0.18 Jy (353 GHz),
0.12 Jy (217 GHz), 0.13 Jy (143 GHz), 0.16 Jy (100 GHz HFI), 0.19 Jy
(100 GHz LFI), 0.24 Jy (70 GHz), 0.25 Jy (44 GHz) and 0.23 Jy (30
GHz). We detect around 27700 PS at the highest frequency
Planck channel and 2900 at the 30 GHz one. The completeness level are:
70\% (857 GHz), 75\% (545 GHz), 70\% (353 GHz), 80\% (217 GHz), 
90\% (143 GHz), 85\% (100 GHz HFI), 80\% (100 GHz LFI), 
80\% (70 GHz), 85\% (44 GHz) and 80\% (30 GHz).
In addition, we can find
several PS at different channels, allowing the study of the spectral
behaviour and the physical processes acting on them.
We also present the basic procedure to apply the method in maps convolved
with asymmetric beams.
The algorithm takes $\sim$ 72
hours for the most CPU time demanding channel
(857 GHz) in a Compaq HPC320 (Alpha EV68 1 GHz
processor) and requires 4 GB of RAM memory; the CPU time  goes as 
$O(N_{R_o}{N_{\rm pix}}^{3/2}log(N_{\rm pix}))$, 
where $N_{\rm pix}$ is the number of pixels in
the map and $N_{R_o}$ is the number of optimal scales needed.
\end{abstract}


\section{Introduction}
\label{intro}
\footnotetext{e-mail: vielva@ifca.unican.es}
The study of the Cosmic Microwave Background (CMB) anisotropies
is one of the most powerful tools to understand the Universe. 
The CMB power spectrum analysis provides us information about how
the Universe was as early as 300000 years after the Big Bang.
Acoustic peaks were predicted to be present in the CMB power spectrum
for several cosmological models (see Hu et al. 1997 for a review).
Recent experiments like BOOMERanG (Netterfield et al. 2002, Rulh et
al. 2002), MAXIMA
(Hanany et al. 2000), DASI (Halverson et al. 2002), VSA
(Rubi{\~n}o-Mart{\'\i}n et al. 2002), CBI (Mason et al. 2002), ACBAR 
(Kuo et al. 2002) and Archeops (Benoit et al. 2003) have
shown the presence of peaks predicted for those flat models.
Moreover, very recently the first detection of the E-mode polarisation
in the CMB has been claimed (DASI, Kovac et al. 2002). That detection
gives even more support to structure formation models via
gravitational instability.
However, there are several cosmological parameters that have not been
estimated yet, due to the tight experimental requirements needed
to put constraints on them.
Two ambitious projects have been proposed to measure the CMB
anisotropies with enough resolution and sensitivity.
One of them is the NASA WMAP mission (Bennet et al. 1996), that was launched
in 2001 and which first year data have been released (Bennett et al. 2003
and references therein)

The second one is the most important CMB experiment developed
up to date: the ESA Planck mission; it will be launched in 2007
and will provide 10 all-sky maps at 9 different frequencies.
Planck mission has two different instruments:
the Low Frequency Instrument (LFI, Mandolesi et al. 1998) and
the High Frequency Instrument (HFI, Puget et al. 1998). The LFI
has a set of radiometers at 30, 44, 70 and 100 GHz; whereas the
HFI has bolometers at 100, 143, 217, 353, 545 and 857 GHz.
The resolution goes from 5$'$ at high
frequencies to 33$'$ at the lowest one. The sensitivity goes from a few
$\mu$K at low frequencies to $\sim 10$mK at the highest
frequency channel. In addition, Planck will generate polarisation maps with
good resolution and sensitivity. Thanks to all these properties,
Planck data will put constraints for all the cosmological
parameters with errors lower than $1\%$.

Although the Planck frequency range is selected to have a low contribution
from additional sources, there are several foregrounds that have
an important emission within this frequency range.
The cleaning up of the microwave
maps is one of the most important challenges in the
CMB data analysis. It is necessary --in order to have CMB
maps as accurate as possible-- to apply mathematical tools to
remove foreground contribution in the microwave sky.
Moreover, the better knowledge of these foregrounds is another
important goal of the Planck mission.
There are several features of the foregrounds (Galactic
emissions and extra-galactic sources) that are poorly known
(like spatial distribution, spectral indices, new extra-galactic
source populations, dust temperature, etc.). With
the data provided by Planck, we expect to understand some of them. 

During the last years, a large number of mathematical tools have been
proposed to perform the \emph{component separation}: Wiener filter
(Tegmark \& Efstathiou 1996, Bouchet \& Gispert 1999), Maximum Entropy
Methods (MEM, Hobson et al. 1998, Stolyarov et al. 2001),
Independent Component Analysis (Baccigalupi et al. 2000,
Maino et al. 2002). All these works have achieved good
component separations, not only on simulated maps, but also on
real data (Barreiro et al. 2003).
However, there are some problems with these all-component separation
methods. The most general one is related to the philosophy of the
methods: the sum of all the recovered components must be equal to the analysed
map; in other words, if a component has not been recovered correctly,
it is possible that another one has been affected by the unrecovered
signal. Fortunately, due to the frequency range that is scanned,
the cosmological signal is recovered with good precision, albeit
some Galactic emissions are poorly recovered.
The second problem we would like to point out is that some components
are not well described by the particular assumptions made by the method.
All these techniques
assume that each component emission can be factorized
in both a spatial template and a frequency dependent function. This is only
true for the CMB emission and the thermal Sunyaev-Zel'dovich effect
(SZ). It is not a bad approximation for the Galactic emission (at least
over small and medium size areas). However, it is a really bad approximation for
the point source emission, and thus other alternatives must be considered
to deal with the point source emission (like modeling it as a noise
contribution, Hobson et al. 1999) . In fact, the point source emission is the most
problematic one for the all-component separation methods and their
contribution to the CMB power spectrum could be important at small
scales (Toffolatti et al. 1998).

In order to avoid these problems, other methods have been
proposed to remove just one of the
microwave sky components. The maximum effort have been done to
subtract the emission due to extra-galactic point sources:
matched filter (Tegmark \& Oliveira-Costa
1998, Naselsky et al. 2002),
scale-adaptive filter (SAF, Sanz et al. 2001, Herranz et al. 2002a,
Chiang et al. 2002) and
based on the Mexican Hat Wavelet techniques
(MHW, Cay{\'o}n et al. 2000, Vielva et al. 2001a).
Recently, some papers have appeared focused on the
SZ detection: SAF (Herranz et al. 2002b, 2002c) and Bayesian approaches
(Diego et al. 2002). The results achieved with these methods are promising,
since the number of detections and the recovery errors are very good.
In addition, the assumptions made in the analysis are very
simple.
Looking at the previous facts, it is obvious that
a combination of both kind of methods (all-component
and one-component separation) can achieve better results.
For example, the combination of MEM and the MHW
(Vielva et al. 2001b) have obtained a very good separation.

\begin{table}
%
%
   \begin{center}
         \begin{tabular}{|c|c|c|c|c|}
	 \hline
	 Frequency & FWHM & Pixel size & $N_{\rm side}$ & $\sigma_{noise}$ \\
	 (GHz) & (arcmin) & (arcmin) &  & $(10^{-6})$ \\
	 \hline
	 857 & 5.0 & 1.72 & 2048 & 19370.15 \\
	 \hline
	 545 & 5.0 & 1.72 & 2048 & 426.90 \\
	 \hline
	 353 & 5.0 & 1.72 & 2048 & 41.82 \\
	 \hline
	 217 & 5.5 & 1.72 & 2048 & 13.76 \\
	 \hline
	 143 & 8.0 & 3.44 & 1024 & 4.65 \\
	 \hline
	 100 (HFI) & 10.7 & 3.44 & 1024 & 5.29 \\
	 \hline
	 100 (LFI) & 10.0 & 3.44 & 1024 & 12.49 \\
	 \hline
	 70 & 14.0 & 3.44 & 1024 & 14.66 \\
	 \hline
	 44 & 23.0 & 6.87 & 512 & 5.93 \\
	 \hline
	 30 & 33.0 & 13.74 & 256 & 3.84 \\
	 \hline
      \end{tabular}
      \caption{\label{planck}
	Experimental constrains at the 10 Planck channels. The
      antenna FWHM is given in column 2 for the different frequencies
      (a Gaussian pattern is assumed).
      Characteristic pixel sizes are shown in column 3. We show the
      \emph{$N_{\rm side}$} HEALPix parameter in column 4. The
      fifth column contains information about the instrumental noise
      level, in $\Delta T/T$ per pixel.}
    \end{center}
\end{table}
In this paper we extend the work in Vielva et
al. (2001a). In that paper, the MHW technique was applied to detect point
sources in simulated flat sky patches. Now, we apply the spherical
generalization of the MHW to detect the point source emission
in all-sky maps, following the Planck mission characteristics.
We present a Planck point source catalogue that could be
achieved with the Spherical Mexican Hat Wavelet (SMHW,
Mart{\'\i}nez-Gonz{\'a}lez et al. 2002). The HEALPix scheme
(G{\'o}rski et al. 1999) is used since it is the
one expected for the Planck data.

\begin{table*}
%
%
   \begin{center}
         \begin{tabular}{|c|c|c|c|c|c|c|c|c|}
	 \hline
	 Frequency & CMB & SZ & 
         TDust     & TDust             & 
         Free-free & Free-free         & 
         Synch.    &  Synch.             \\
	 (GHz)     & ($\times10^{-5}$) & ($\times10^{-6}$)   &
         all-sky   & $|b|>50^{\circ}$  & 
         all-sky   & $|b|>50^{\circ}$  &
         all-sky   & $|b|>50^{\circ}$  \\
	 \hline
	 857       & $4.29$ & $30.10$ &
                     $12.56$             & $0.14$         & 
                     $1.75\times10^{-3}$ & $1.32\times10^{-5}$ & 
                     $3.07\times10^{-4}$ & $6.79\times10^{-5}$  \\
	 \hline
	 545       & $4.29$ & $15.2$ & 
                     $9.40\times10^{-2}$ & $1.12\times10^{-3}$ &
                     $4.68\times10^{-5}$ & $3.75\times10^{-7}$ &
                     $1.03\times10^{-5}$ & $2.14\times10^{-6}$  \\
	 \hline
	 353       & $4.29$ & $6.04$ &
                     $4.87\times10^{-3}$ & $6.06\times10^{-5}$ &
                     $9.59\times10^{-6}$ & $8.02\times10^{-8}$ &
                     $2.60\times10^{-6}$ & $5.08\times10^{-7}$  \\ 
	 \hline
	 217       & $4.29$ & $0$                 &
                     $6.12\times10^{-4}$ & $7.87\times10^{-6}$ & 
                     $6.31\times10^{-6}$ & $5.46\times10^{-8}$ & 
                     $2.18\times10^{-6}$ & $3.93\times10^{-7}$  \\ 
	 \hline
	 143       & $4.26$ & $2.36$ &
                     $1.88\times10^{-4}$ & $2.46\times10^{-6}$ &
                     $8.53\times10^{-6}$ & $7.53\times10^{-8}$ &
                     $3.16\times10^{-6}$ & $6.07\times10^{-7}$ \\ 
	 \hline
	 100       & $4.26$ & $3.41$ &
                     $8.55\times10^{-5}$ & $1.13\times10^{-6}$ & 
                     $1.43\times10^{-5}$ & $1.28\times10^{-7}$ &
                     $7.18\times10^{-6}$ & $1.14\times10^{-6}$  \\ 
	 (HFI, LFI) & & & & & & & & \\ 
	 \hline
	 70        & $4.26$ & $3.96$ & 
                     $4.33\times10^{-5}$ & $5.78\times10^{-7}$ &
                     $2.72\times10^{-5}$ & $2.44\times10^{-7}$ &
                     $1.61\times10^{-5}$ & $2.42\times10^{-7}$ \\ 
	 \hline
	 44        & $4.17$ & $3.38$ &
                     $1.92\times10^{-5}$ & $2.57\times10^{-7}$ & 
                     $6.87\times10^{-5}$ & $6.21\times10^{-7}$ & 
                     $5.05\times10^{-5}$ & $7.05\times10^{-6}$  \\ 
	 \hline
	 30        & $3.99$ & $2.55$ &
                     $1.01\times10^{-5}$ & $1.36\times10^{-7}$ &
                     $1.53\times10^{-4}$ & $1.39\times10^{-6}$ &
                     $1.35\times10^{-4}$ & $1.77\times10^{-5}$   \\ 
	 \hline
      \end{tabular}
      \caption{\label{rms}
		We show the RMS values for the map components at the
               Planck channels. The data correspond to the unconvolved
               maps are given in $\Delta T/T$ units. The CMB RMS
               are presented in the second column; as it is well known,
               the CMB is frequency independent: the small increase
               from low to high frequency is due to the different
               pixelization. In the third column the SZ RMS are shown.
	       The thermal dust, free-free and synchrotron all-sky
               RMSs values are presented in columns 4, 6 and 8 respectively;
	       whereas in columns 5, 7 and 9 we report the RMS for
               the same components at Galactic latitudes (absolute value)
               above $50^{\circ}$. As the main results of this work
               are presented for simulations that do not include rotational
               dust emission, we do not include in the Table the
               RMS values. We just describe qualitatively its importance:
               it is the dominant all-sky emission from 30 GHz 
               ($\approx 10$ times larger)
               up to 70 GHz ($\approx 2$ times larger),
               whereas at Galactic latitudes greater than $50^{\circ}$ its
               emission is stronger than the one due to free-free but
               lower than the synchrotron one.}
    \end{center}
\end{table*}
The paper is organized as follows. In Section \ref{simul} we explain
how the simulations have been done and which are the Planck
instrumental features. The SMHW is present in Section \ref{method}
as well as the algorithm and the HEALPix implications to the method.
The results are given in Section \ref{results}. In Section \ref{beams}
we introduce the basic procedure to apply the method in maps convolved
with asymmetric beams.
Finally, the conclusions of this work are presented in Section \ref{fin}.

\section{Simulated all-sky Planck maps}
\begin{figure*}
%
%
	\begin{center}
%
%

%
%
		\includegraphics[width=16cm]{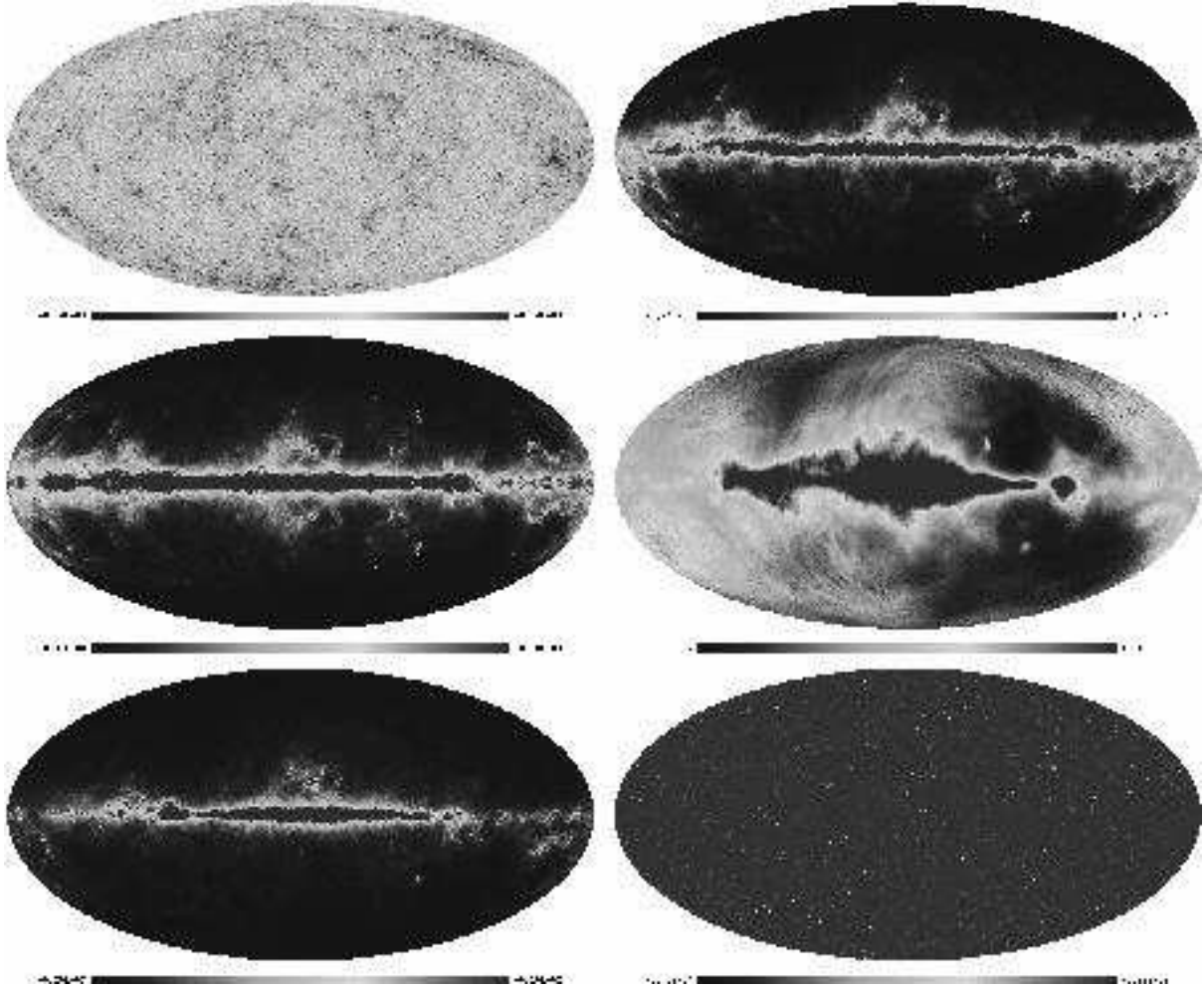}

\caption{\label{compo}
Simulated components at 30 GHz. We show (from
up-left to bottom-right,) CMB, thermal dust, free-free,
synchrotron, rotational dust and SZ. The maps are in
$\Delta T/T$ thermodynamic temperature units.}
\end{center}
\end{figure*}
\begin{figure*}
%
%
	\begin{center}
%
%
%

%
		\includegraphics[width=16cm]{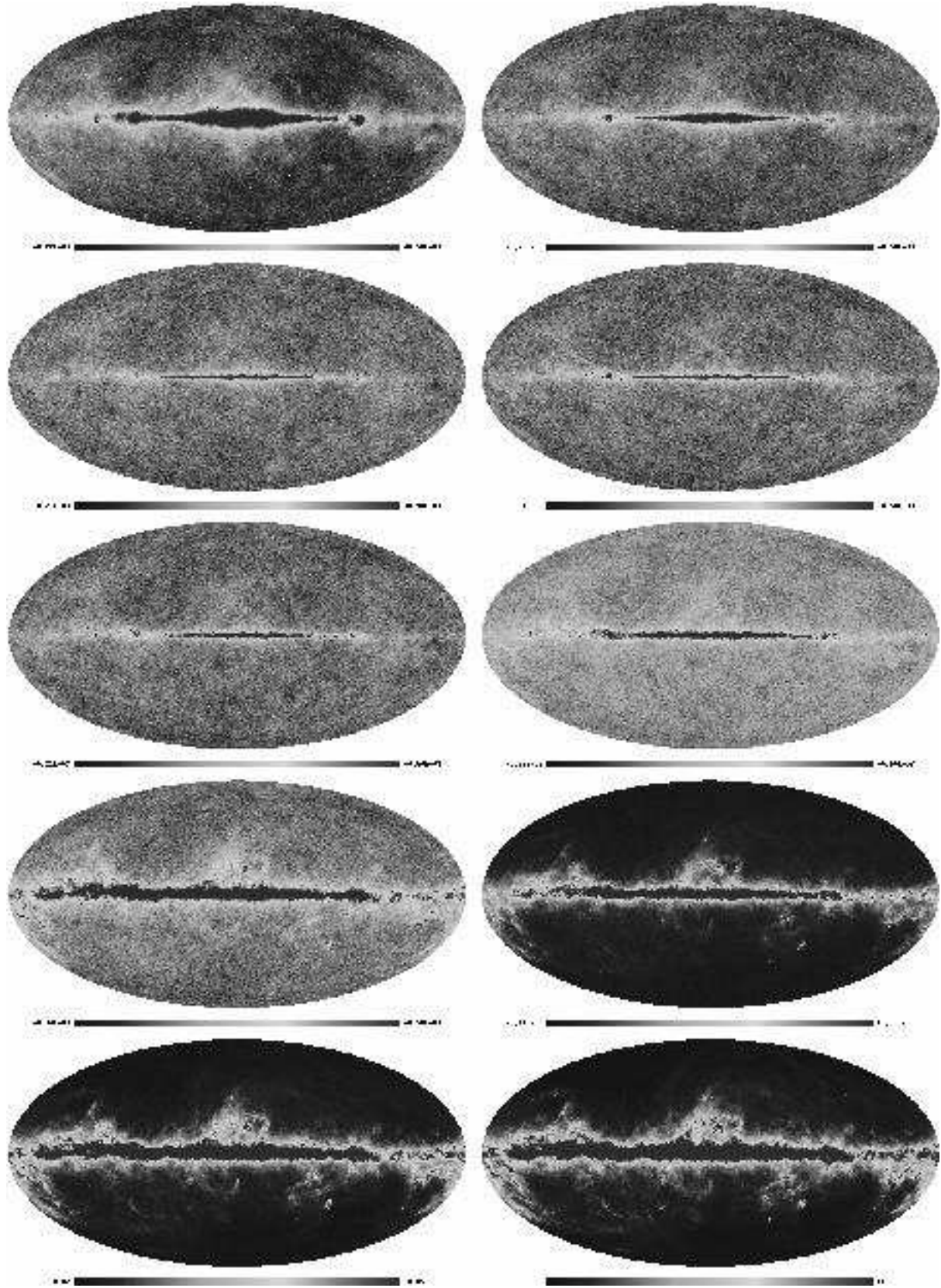}

\caption{\label{skyMAPS}
Simulated all-sky Planck maps. The frequencies are (from
up-left to bottom-right): 30, 44, 70, 100(LFI), 100(HFI), 143, 217,
353, 545 and 857 GHz. All the maps are filtered with a Gaussian beam
and noise have been added following the RMS levels of Table \ref{planck}.
They are in $\Delta T/T$ thermodynamic temperature units.}
\end{center}
\end{figure*}

\label{simul}
In order to test the capabilities of the SMHW to detect point sources,
we have simulated a data set like the one expected from the Planck
mission. We have ten all-sky maps --one for each Planck channel-- at nine
different frequencies. In Table \ref{planck} we show the pixel
sizes, antenna FWHMs, HEALPix $N_{\rm side}$ parameters and the expected
Planck noise levels (which is assumed to be Gaussian and pure stationary
white noise) for all the
channels that we have used in the present work. The simulations
include PS emission, CMB, Galactic foregrounds
(thermal dust, free-free, synchrotron) and SZ (see Fig. \ref{compo})
as well as instrumental white noise. In Fig. \ref{skyMAPS} the ten
sky maps are shown, and in Table \ref{rms}
the RMS values for all the components are presented.

The foreground due to the thermal dust emission have been
simulated using the data and the model provided by
Finkbeiner et al. (1999). That emission is modeled by
two gray-bodies with different temperatures from
point to point: a \emph{hot} one with a mean
dust temperature of ${T_D}^{hot} \simeq 16.2$K and an
emissivity $\alpha^{hot} \simeq 2.70$, and a \emph{cold} one with a
mean ${T_D}^{cold} \simeq 9.4$K, and $\alpha^{cold} \simeq 1.67$.
Free-free emission is poorly known.
Present experiments such as SouthernH-$\alpha$ Sky Survey
(SHASSA, Reynolds \& Haffner, 2000) and the Wisconsin Halpha Mapper
project (WHAM, Gaustad et al. 2001) 
will provide maps of $H_{\alpha}$ emission that
could be used as a template for this emission.
At the moment this work was done, free-free maps
were not available. We have chosen the idea proposed
by Stolyarov et al. (2002) to simulate this component
assuming that a $60\%$ of the signal is a thermal dust correlated
component,
whereas the rest of the emission is uncorrelated (to simulated this
uncorrelated component, the flipped thermal dust map is used as template).
Synchrotron emission simulations have been done using the
all-sky templates given by Giardino et al. (2002).
These maps
are an extrapolation of the 408 MHz radio map of Haslam et al.
(1982), from the original $1^{\circ}$ resolution.
A power law for the power spectrum 
with an exponent of $-3$ has been assumed. 
We include in our simulations the information on the 
spectral index variation as a function of electron density in the Galaxy.
This spectral indices template have been done combining the 408 MHz
map with the Jonas et al. (1998) one at 2326 MHz and the Reich
\& Reich (1986) map at 1420 MHz and was done also by
Giardino et al. (2002).
Although the results of this paper are given for simulated sky maps
where only the previous Galactic emissions are included, we have
tested how the presence of rotational dust (Draine \& Lazarian 1998)
could modify the PS catalogue.
Rotational dust emission could be
important at the lowest frequencies
(30 and 44 GHz), where it could be
around ten times greater than the all-sky free-free emission.
This emission is 
strongly correlated with the thermal dust one,
through the neutral hydrogen column density ($N_H$):
\begin{equation}
{I(\nu)}_{\textrm{rot}} = f(\nu)N_H, \hspace{0.5cm}
{I(3000~{\rm GHz})_{\textrm{thermal}}} = a N_H,
\nonumber
\end{equation}
where $f(\nu)$ is the frequency dependence of the emissivity
predicted by Draine \& Lazarian (1998)
and $a$ is the correlation between the $21$cm emission and the
infrared dust one. We adopt the correlation proposed by Boulanger
\& P{\'e}rault (1988):
\begin{equation}
a \approx 0.85\times 10^{-14} {\rm Jy~ sr}^{-1}
{\Big(\frac{H~\textrm{atoms}}{\textrm{cm}^{-2}}\Big)}^{-1}.
\nonumber
\end{equation}
Therefore, the rotational dust emission is simulated
from the thermal one through the equation:
\begin{equation}
{I(\nu)}_{\textrm{rot}} =  a^{-1} f(\nu) 
{I(3000~{\rm GHz})_{\textrm{thermal}}}.
\nonumber
\end{equation}
The Sunyaev-Zel'dovich (SZ) effect emission have been developed
following the model proposed by Diego et al. (2001).
These simulations assume a flat $\Lambda$CDM Universe with
$\Omega_m = 0.3$ and $\Omega_{\Lambda}$ = $0.7$.
The CMB signal have been simulated for the same Universe,
using the $C_\ell$ generated with the CMBFAST
code (Seljak \& Zaldarriaga, 1996).
Finally, the extra-galactic point source (PS) simulations have been 
performed following the model of Toffolatti et al. (1998)
assuming the above adopted cosmological model. Two main PS
populations are assumed. 
At low and intermediate frequencies (from 30 to $\sim 300$ GHz)
flat spectrum sources (QSOs, blazars and AGN) dominate bright
source counts (i.e., at $S\sim$ 50$\div$100 mJy). This is a model
prediction, given that only few data are currently available in
this frequency range. On the other hand, the model counts of
Toffolatti et al. have been very precisely confirmed
by independent observations (CBI, Mason et al., 2002; VSA, Taylor
et al., 2002) and, moreover, by the full sky sample of
extragalactic sources released by the WMAP satellite (Bennett et al.,
2003). At frequencies $\nu \sim 300$ GHz, number counts of extragalactic 
sources are dominated by dusty galaxies. High redshift spheroids
and elliptical galaxies in the phase of rapid star formation and
low redshift starburst and normal spiral galaxies. The model number
counts of Toffolatti et al. (1998) are still assumed, albeit the new
data by SCUBA and MAMBO surveys at sub--mm/mm wavelengths are 
indicating a greater slope of differential counts at fluxes of a few
mJy.
\section{The method}
\label{method}
The proposed PS detection method is composed of several pieces. Its basic pillar
is the Spherical Mexican Hat Wavelet (SMHW), that is used to convolved
the analysed signal at different scales in order to achieve a \emph{cleaned}
map; we describe the SMHW in Subsection \ref{SMHW}. Because the
expected pixelization for the Planck maps is the HEALPix scheme, we
have developed the method in this framework; this implies some
peculiar characteristics that will be discussed in
Subsection \ref{healpix}. The proposed method is not simply a convolution
of the maps with the SMHW, there are several steps in the algorithm that
are commented in Subsection \ref{algorithm}. One of the important
steps in the PS catalogue estimation is the detection criterion:
we will discuss about this topic in Subsection \ref{detect}.
\subsection{The tool: Spherical Mexican Hat Wavelet}
\begin{figure}
%
%
	\begin{center}
		\includegraphics[width=8cm]{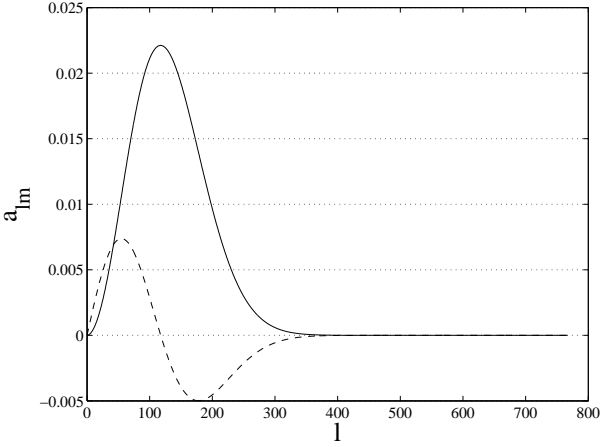}

\caption{\label{alms}
We show the SMHW $a_{lm}$ for $R = 42$ arcmin (the largest scale
used in this work). In addition, we plot (dashed line) the difference
between the SMHW $a_{lm}$ and the MHW one, 50 times magnified.}
\end{center}
\end{figure}
\label{SMHW}
The SMHW has been used recently to study the CMB
Gaussianity/Non-Gaussianity in the COBE-DMR data (Cay{\'o}n et al. 2002)
and in Planck simulations (Mart{\'\i}nez-Gonz{\'a}lez et al. 2002).
The extension of the plain Mexican Hat Wavelet (MHW) to the sphere 
(like the spherical extension of any wavelet) is not an obvious issue.
The stereographic projection has been suggested by Antoine \&
Vanderheynst (1998) as the most suitable manner to do this
extension, since the MHW properties are kept and at the
small angle limit tends to the MHW. A graphical explanation of such
extension can be found in Mart{\'\i}nez-Gonz{\'a}lez et al. (2002), here we just
remark the expression for the MHW and the SMHW.
The MHW is given by Eq. (\ref{eqMHW})
and it satisfies the \emph{compensation} ($\int d\vec{x}\,\Psi = 0$),
\emph{admissibility}
($C_{\Psi} = (2\pi )^2\int_0^{\infty}dk\,k^{-1}{\Psi}^2(k) < \infty$,
where $\Psi (k)$ is the Fourier transform of $\Psi (x)$)
and \emph{normalization} ($\int d\vec{x}\,{\Psi}^2 = 1$)
properties that define a wavelet (see
Mart{\'\i}nez-Gonz{\'a}lez et al. 2002 for details):
\begin{equation}
\label{eqMHW}
   \Psi(x) = \frac{1}{\sqrt{2\pi}}\frac{1}{R}\Big[2-\big(\frac{x}{R}\big)^2\Big]
             e^{-\frac{x^2}{2R^2}},
\end{equation}
in that expression $R$ is the MHW scale and $x$ is the
distance. Through the stereographic projection, we deal with the SMHW
expression:
\begin{equation}
\label{eqSMHW}
   \Psi_S(y,R) = \frac{1}{\sqrt{2\pi}N(R)}{\Big[1+{\big(\frac{y}{2}\big)}^2\Big]}^2
  \Big[2 - {\big(\frac{y}{R}\big)}^2\Big]e^{-{y}^2/2R^2},
\end{equation}
where $R$ is the scale and $N(R)$ is a normalization constant:
\begin{equation}
\label{N}
      N(R)\equiv R{\Big(1 + \frac{R^2}{2} + \frac{R^4}{4}\Big)}^{1/2}.
\end{equation}
The distance on the tangent plane is given by $y$ that is related 
to the latitude angle ($\theta$) through:
\begin{equation}
\label{y}
      y\equiv 2\tan \frac{\theta}{2}.
\end{equation}
Given the SMHW, a signal on the sky $f(\vec n)$ (being $\vec{n}$
a vector in the projection plane) can be analysed,
obtaining the wavelet coefficients:
\begin{equation}
\label{coeff}
	w(\vec n, R) = \int d\Omega' \, f(\vec n + \vec n')
	\Psi_S(\theta', R).
\end{equation}
where
\begin{eqnarray}
\vec{n} & \equiv  & 2\tan \frac{\theta}{2}(\cos \phi , \sin \phi ), \\
\vec{n'} & \equiv & 2\tan \frac{\theta^{\prime}}{2}(\cos \phi ' , \sin
\phi')
\end{eqnarray}
If the signal $f(\vec n)$ is a PS with a Gaussian shape, then the
wavelet coefficient at the PS position is a known function of
the antenna width ($\sigma_a$), the SMHW scale ($R$) an
the PS intensity ($I$). 

For the SMHW scales used in this work (smaller than $1^\circ$) the SMHW
can be described by the plain MHW with good accuracy. In
Fig. \ref{alms} we have plotted (solid line) the spherical harmonic
coefficients ($a_{lm}$) of the
SMHW at the largest scales used in our work ($R = 42$ arcmin); we also
plot (50 times magnified) the difference between
these coefficients and the Fourier
coefficients that describe the MHW at the same scale (dashed
line). This implies that, for the typical cases studied in this work,
we consider the MHW and the SMHW coefficients to be the same
for all practical purposes.
Therefore, for the optimal scale determination through the
plane projection (see next Section) and for the analytical expression
relating the wavelet coefficient at the scale $R$ 
with the beam dispersion ($\sigma_a$)
and the point source amplitude ($I$) (see Eq.~\ref{expression}), the
MHW is used. On the other hand, all-sky convolutions are performed
using the SMHW.
\begin{equation}
\label{expression}
w(R) = 2\sqrt{2\pi}IR\frac{(R/\sigma_a)^2}{(1 +
(R/\sigma_a)^2)^2}.
\end{equation}

We are interested in detecting PS in wavelet space rather than real
space because, by convolving the map with the SMHW,
we can increase the signal-to-noise ratio of the sources. This increment
is characterised by the \emph{amplification} factor:
\begin{equation}
\label{amplification}
A = \frac{w(R)/\sigma_w(R)}{I/\sigma_m},
\end{equation}
where $\sigma_w(R)$ is the dispersion of the wavelet coefficients
and $\sigma_m$ is the dispersion of the analysed signal.
If we assume that the background is an statistically
homogeneous and isotropic random field with zero mean,
and taking into account that its variance is given by:
\begin{equation}
{\sigma^2}_{w}(R) = <w(\vec{n}, R)^2> - <w(\vec{n}, R)>^2,
\end{equation}
then it is straightforward to show that:
\begin{equation}
\label{sigma}
{{\sigma}^2}_{w}(R) \propto \int d\,q q \, P(q) |\tilde{\Psi}(qR)|^2.
\end{equation}
where $P(q)$ is the power spectrum of the analysed signal and
$\tilde{\Psi}(qR)$ is the Fourier transform of the MHW.

The amplification reaches its maximum value at the \emph{optimal
scale}, $R_o$ (see Vielva et al. 2001a for a detailed
description).
Taking into account Eqs. (\ref{expression}), (\ref{amplification}) and
(\ref{sigma}), we can calculate the optimal scale from
the data itself ($C_\ell$, $\sigma_m$ and $\sigma_a$).
Once the PS position is determined (by the location of the maxima in
wavelet space), we can estimate the PS
intensity by calculating the wavelet transform
at each selected location at several scales, in order to 
compare these values with the theoretical curve (Eq. \ref{expression}).
We can define a $\chi^2$ at each selected position $\vec{n}$:
\begin{equation}
\label{chi2}
\chi^2(\vec{n}) = \sum_{i,j}
({w_{\vec{n},R_i}}^{t} - {w_{\vec{n},R_i}}^{e})V_{ij}^{-1}
({w_{\vec{n},R_j}}^{t} - {w_{\vec{n},R_j}}^{e}),
\end{equation}
where $V_{ij}$ represents the wavelet coefficients
covariance matrix element
between scales $i$ and $j$ (calculated from the data)
and ${w_{\vec{n},R_i}}^{t}$ and ${w_{\vec{n},R_i}}^{e}$ are the theoretical
(Eq. \ref{expression}) and experimental values of the wavelet
coefficients at scale $R_i$ and location $\vec{n}$. We
use four different scales to perform the multiscale fit
(see Vielva et al 2001a for a detailed description of
the covariance matrix and the multiscale fit).

\subsection{The framework: HEALPix}
\begin{figure*}
%
%
	\begin{center}
%

		\includegraphics[angle=270,width=16cm]{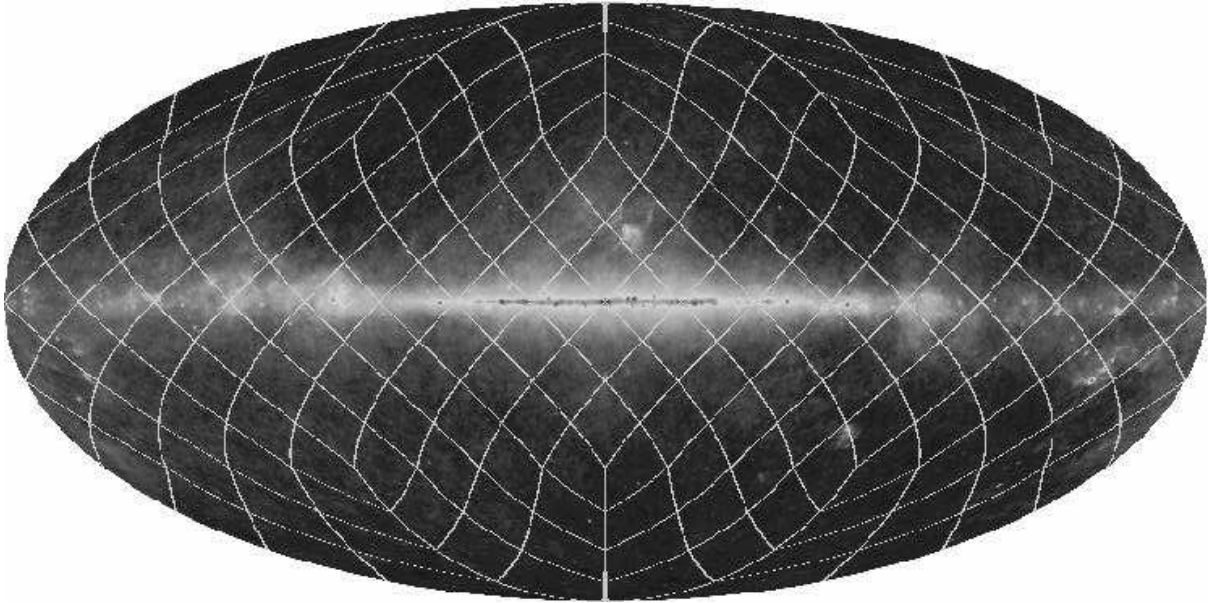}

\caption{\label{esquema}
Sky division in \emph{father pixels} ($N_{\rm side}$ = 4). The optimal
scale is determine in each one of these areas.}
\end{center}
\end{figure*}
\begin{figure*}
%
%
	\begin{center}
%

		\includegraphics[angle=270,width=16cm]{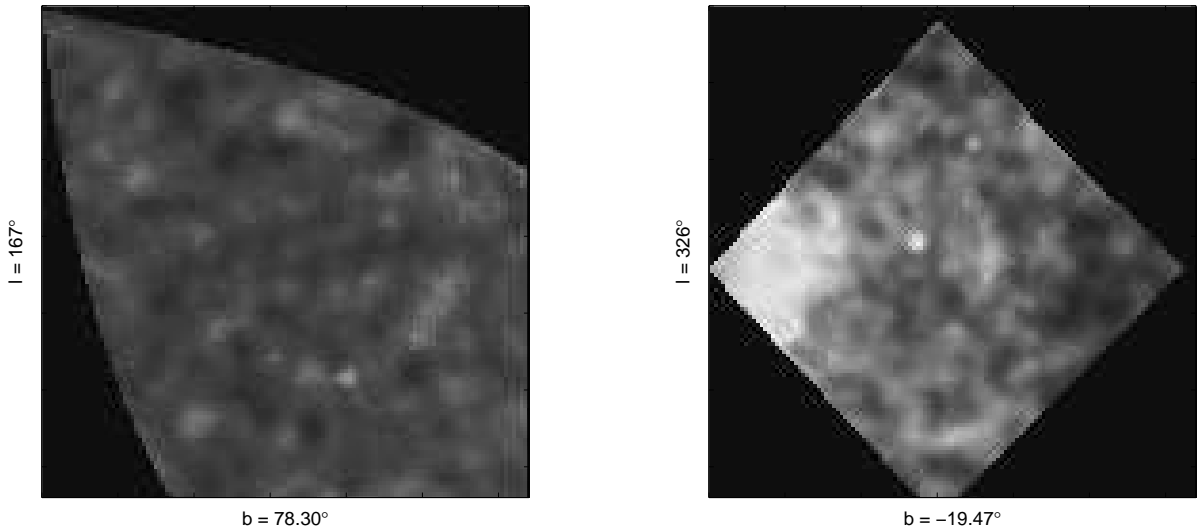}

\caption{\label{projections}
We show the projections of two father pixels in squared plain
patches. The left one represents  one of the four father pixels at the
north pole, whereas the one in the right hand shows a father pixel
near the Galactic plane. The error in the optimal scale determination
in the squared patch due to the projection is less than $5\%$.}
\end{center}
\end{figure*}

\label{healpix}
The Hierarchical Equal Area and iso-Latitude Pixelization (HEALPix)
have been suggested by Gorski et al. (1999) as the most suitable
sphere pixelization since it allows 
hierarchical tree structure and fast spherical harmonics transform.
These two properties give us great advantages in order to deal with
the PS detection.
First of all, the hierarchical structure allows
to identify quickly pixels in the sphere, what is specially relevant
for large data sets. We have used this characteristic to determine
the \emph{optimal scale} in different patches of the sky with 
low CPU time consumption. As it is shown in Vielva et al. (2001a) the 
determination of the optimal scale is essential in order to achieve the
largest amplification. The simulated Planck
maps show strong differences from
one direction to another, for example, at 30 GHz, we can see
synchrotron and free-free structure at the Galactic plane, but
the CMB emission dominates at high Galactic latitude. Hence,
a global optimal scale is, clearly, far from being \emph{optimal}.
We have divided the sky in areas that coincide with the HEALPix
pixels at $N_{\rm side}$ = 4 resolution (see Fig. \ref{esquema}). We have
chosen this resolution because the pixels (hereafter we refer
to these as \emph{father pixels}) have a characteristic
size $\sim 10^\circ$, close to the patch size adopted by Vielva et al.
(2001a). One can think of alternatives to determine the
optimal scale. For example, we have tested to determine it in iso-latitude
bands, but we lose precision in the optimal scale determination,
especially at low Galactic latitude.
To calculate the optimal scale the $C_\ell$ of the signal are needed.
The $C_\ell$ calculation of each father pixel on the sphere is, in practice,
an unrealizable task: there are 192 areas and, 
for each one of them we need to calculate the $a_{lm}$ sets
with good accuracy up to $l_{max} = 3N_{\rm side} - 1$.
To determine the
$a_{lm}$ set of an isolated region of the sky is a difficult issue,
we can fill out with zero values all the pixels outside the
particular zone and to proceed as an all-sky $a_{lm}$ calculation
(only the large scales are mislead, but it is not important since
the relevant information to determine the optimal scale is at the smallest
scales).
Another alternative is to use orthonormal basis for a given
cut sky (Mortlock et al. 2002). To calculate these bases for all the
father pixels is a very slow process (it scales as $O({l_{max}}^6)$
for each father pixels); however, once the calculation is
done, the basis can be stored. Unfortunately, as the authors point out,
the required space to store these bases scales as
$O({l_{max}}^4)$. Hence, this alternative is inadequate for the
Planck resolutions.
To solve this issue, we
project each father pixel in a plain tangent squared patch, filling
with zero
those pixels of the patch outside the projected image
(see Fig. \ref{projections}). Afterwards, we can calculate the optimal
scale through the power spectrum of each particular area using
the Fourier transform.
%
%
This approximation makes the computation of
optimal scale easy and fast, 
but we are making
an error in the optimal scale determination due to the
projection. However, we have tested that the mean error
all over the sky is less than $5\%$, when we compare with the optimal
scale determination through the spherical analysis.
This implies an amplification (flux limit)
which is $\approx 1\%$ lower (greater) than
the one obtained through the right spherical procedure.
\subsection{The algorithm}%
\begin{figure*}
%
%
	\begin{center}

		\includegraphics[height=21cm]{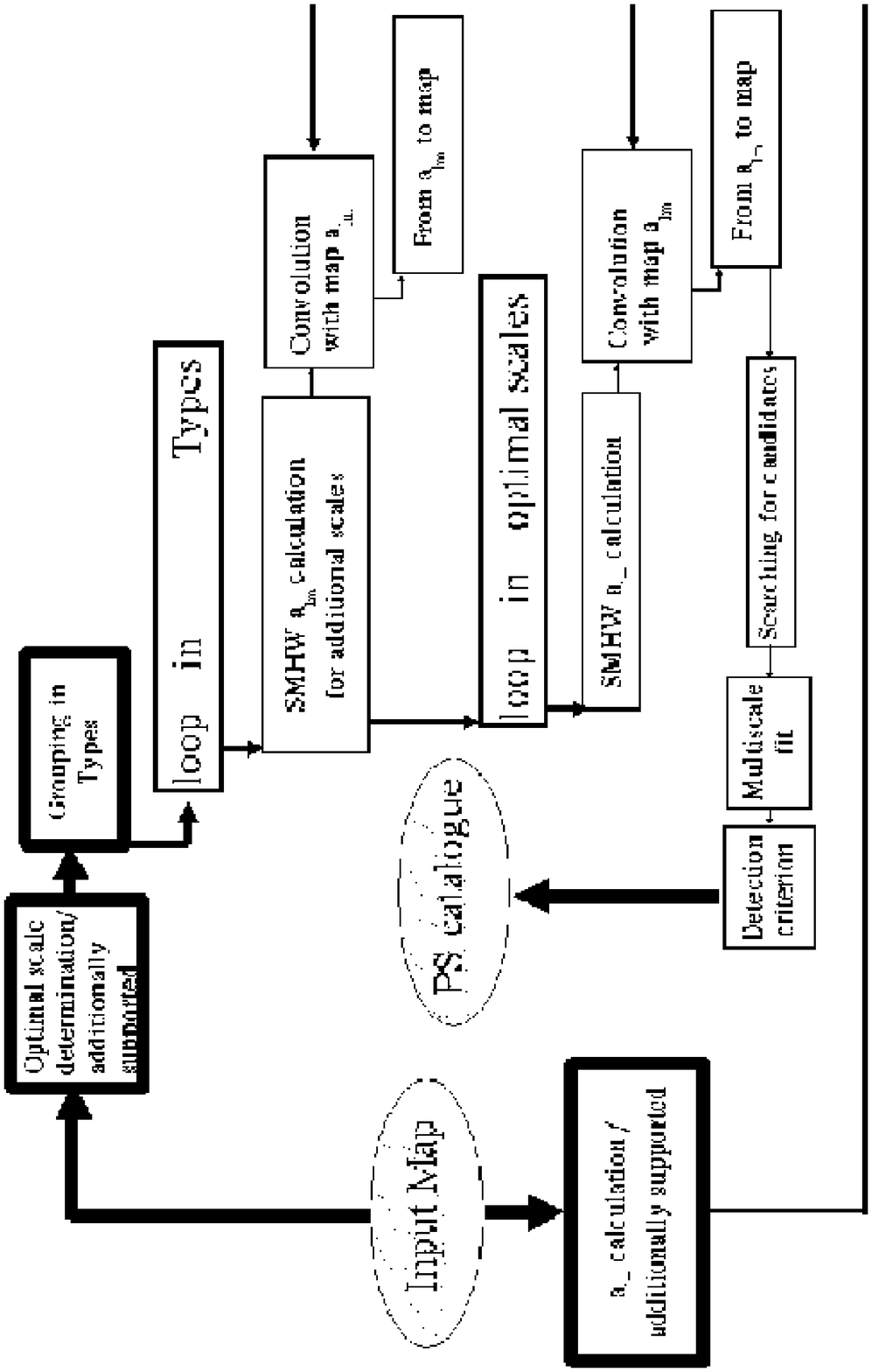}

\caption{\label{algo}
Schematic algorithm of the process. See text for details.}
\end{center}
\end{figure*}

\label{algorithm}

\begin{figure*}
%
%
	\begin{center}

%
%
		\includegraphics[angle=270,width=16cm]{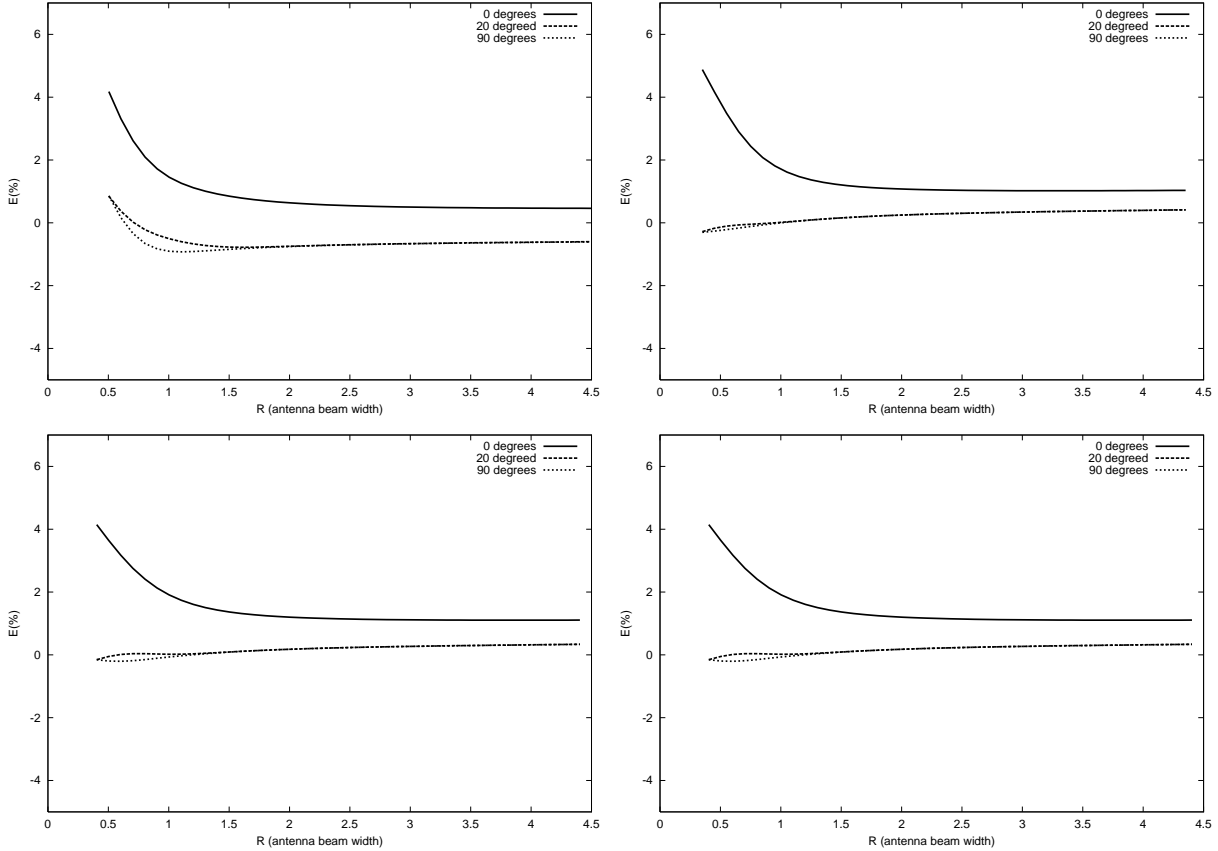}

\caption{\label{RealCoeff}
Deviation of the wavelet coefficients from the theoretical
curve (Eq. \ref{expression}) for several Galactic latitudes. The plots
represent (from up-left to bottom-right) the 44, 100 (LFI),
143 and 545 GHz.}

\end{center}
\end{figure*}

We present the different steps that constitute the PS detection
algorithm. Given a map to be analysed we need to estimate the optimal
scale in each sky area, in the way indicated in the previous
Subsection. Together with the optimal scale, we need other SMHW scales,
in order to perform the multiscale fit. We call these scales
\emph{adjacent scales}. Here we need to make another simplification
in order to reduce the CPU time. If we make the adjacent scale
choice completely dependent on the optimal scale value, we finally
have 4 different SMHW $a_{lm}$ sets for each optimal scale.
We must convolve
the map with each one of these scales and then a transformation
from harmonic space to wavelet space is required
to detect the PS. This last
process (spherical harmonic transform) is the major contribution
to the CPU time consumed. We need to reduce these calculations as
much as possible.
In order to reduce the CPU time, we group the optimal scales
in three \emph{types}. All optimal scales belonging to the same
type have the same adjacent scales. This reduces enormously
the CPU time, making the process viable.
\begin{itemize}
\item type I: the optimal scale is lower than half a pixel. This
              occurs when the dominant background is due to
              foregrounds with a variation scale larger than the PS
	      one (i.e. the beam width) and with a very high emission.
	      This behaviour can be found close to the
              Galactic plane and in some high Galactic emission
              zones. Due to the pixelization, we can not use SMHW
              scales too small, hence a lower limit must be imposed to
              avoid discrepancies between the wavelet coefficients and
              its theoretical value (Eq. \ref{expression}). For that
              reason, we assign the value $R_o \equiv 0.5P_s$ to those
              father pixels belonging to this type I, where 	
              $P_s$ is the pixel size (see Table \ref{planck}).
              In this case, the three adjacent scales
              will be greater than the optimal one.

\item type II: the optimal scale is smaller than the beam width
               but larger than half a
               pixel. This occurs in regions where the major
               background is a signal with a variation scale close to
               the PS one: sky areas dominated by the CMB emission, with a low
               contribution due to instrumental noise and poor resolution
               \footnote{As the typical coherence scale for
               the CMB is $\sim 10'$ and the beam widths are
               larger that this scale (at low
               frequency channels), the effective CMB variation scale
               is close to the antenna dispersion.}.
               This 
               also happens
               for regions with a moderate synchrotron or free-free 
               emission which nonetheless is the dominant one.
	       Most of the optimal scales used in this
               work belong to this type.
 	       In this case the adjacent scales are also greater
	       than the optimal one, but the $R_o$ has not a minimum value.

\item type III: the optimal scale is close to the beam width.
	        The signal components with variation scales larger
	        than the PS one are dominant (Galaxy components),
                but their emission is lower than in previous types.
                In addition, the instrumental noise dominates over the CMB
                emission.
                This happens at very high
	        Galactic latitudes in dust-dominated channels. One of
	        the adjacent scales is smaller than $R_o$, whereas the
	        other two are greater.
\end{itemize}
The wavelet coefficient at the scales that come
from the previous division are well described by
the theoretical curve (Eq. \ref{expression}). Only those coefficients
at scales equal or lower than the pixel size and at high latitude
(where the HEALPix pixels have an elongated shape) are slightly affected by
pixelization problems. However, as Fig.~\ref{RealCoeff} shows, the
errors for most of the cases are less than $5\%$ and close to
zero for the rest of the situations (intermediate and low latitudes and
scales larger than the pixel size).
Once the $a_{lm}$ sets at the adjacent scales are calculated,
the SMHW convolution of the map is performed for these
scales. We want to remark that the SMHW convolution is performed
in all the sky, not only on those father pixels that belong
to the same type: it is a spherical convolution.
We anti-transform the convolved $a_{lm}$ to get the wavelet
coefficients of the sky maps.
To do that efficiently, we use another of the HEALPix
properties: the iso-latitude pixelization.
A map in HEALPix has 4$N_{\rm side}$ - 1 iso-latitude rings,
for each one of these rings, we calculate the Legendre polynomials
(that only depend on the declination angle). Using these
functions we can calculate the harmonic transform of
each ring at the same time for the three adjacent scales.
Afterwards,
we calculate the SMHW $a_{lm}$  sets of each optimal scale of the peculiar
type. We convolve the map and anti-transform the convolved $a_{lm}$
set to the wavelet space and search for maxima in the
optimal scale wavelet coefficients map on these father
pixels belonging to the specific optimal scale.
At this point, we want to comment about a problem related with
the estimation of the mean value of the wavelet coefficients maps.
The maps that
we are analysing have zero mean and  the wavelet transform keeps
the mean value. However, because we are searching for maxima
only on those father pixels with a given optimal scale, the wavelet
coefficients set on those father pixels have not zero mean (the
zero mean is a global property). In other words, the zero level
is mislead. This is an important problem at high frequency channels
(from 353 GHz to 857 GHz), where the Galactic contribution
is so high that the mean is completely dominated by it: the SMHW convolution
is not able to remove sufficiently the Galactic contribution
(near the Galactic plane).
It implies an offset in the wavelet coefficients and
also in the PS amplitudes.
To solve this
problem, we recalculate the zero level of each wavelet coefficients
set. The bias in the amplitude estimation is not removed completely;
however, as it can be seen in the next
Section it is not important and it can be determined.
Once the maxima are located we perform a multiscale fit using the
four scales (the optimal and the 3 adjacent ones) to estimate
the PS amplitude.
We repeat this algorithm for all the optimal scales in each type
and for all the types. In Fig \ref{algo} we show an schematic
diagram of the algorithm.
The process takes 72 hours for the \emph{worse} case: the 857 GHz
map, $N_{\rm side}$ = 2048.
We have run the code in a Compaq HPC320 (Alpha EV68 1 GHz
processor) and it requires 4 GB of RAM memory. The $a_{lm}$ map
estimation takes $\sim$ 8 hours, since we need to have scale
information up to the highest resolution $l = 3N_{\rm side} - 1$.
The optimal scale determination is really
fast ($\sim 15$ minutes). The rest of the time is due to
transformations
from harmonic domain to wavelet space (we have used 17 different
optimal scales for the 857 GHz map). The CPU time goes, basically,
as $O(N_{R_o}{N_{\rm pix}}^{3/2}log(N_{\rm pix}))$,
where $N_{\rm pix}$ is the number of pixels
in the map ($N_{\rm pix} = 12{N_{\rm side}}^2$)
and $N_{R_o}$ is the number of optimal scales. 
\begin{figure*}
%
%
	\begin{center}

%
%
		\includegraphics[width=16cm]{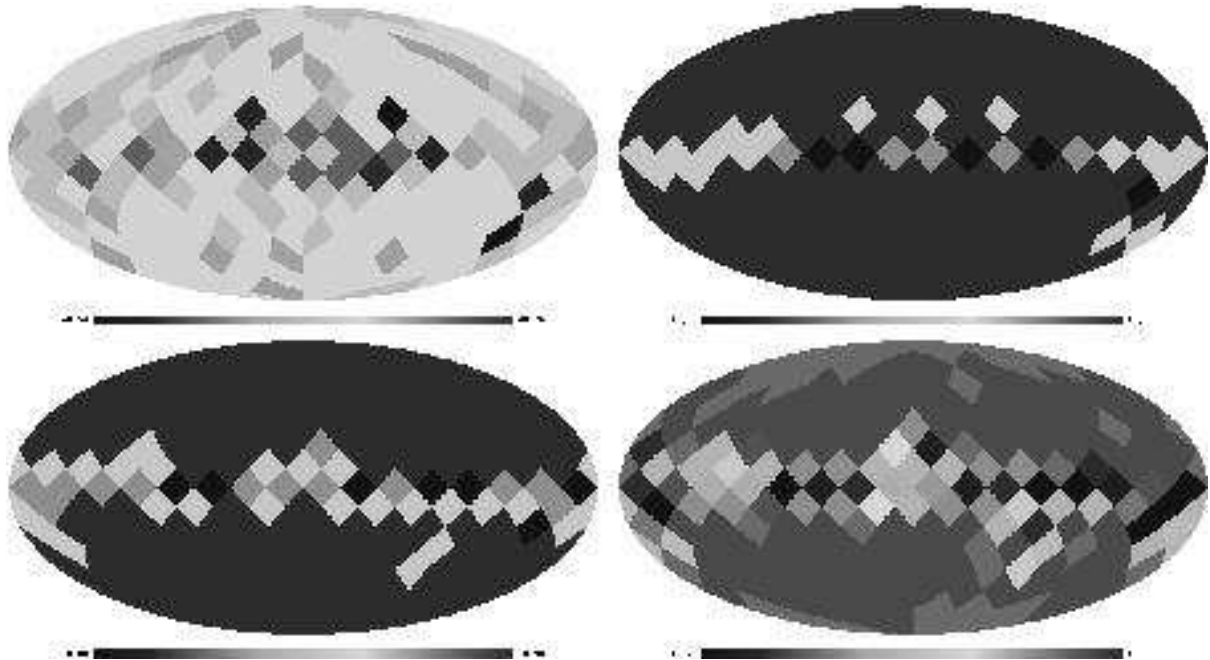}

\caption{\label{OptimalScales}
We present the optimal scale maps of four channels: 44 GHz
(top-left), 100 GHz LFI (top-right), 217 GHz (bottom-left) and 545 GHz
(bottom-right). They are in antenna width ($\sigma_a$)
units (see Table \ref{planck}).}
\end{center}
\end{figure*}
\subsection{The detection criterion}
\label{detect}
To define a detection criterion is an important task in the field
of the PS detection. A good detection criterion should be
robust, efficient and unbiased. The robustness property implies
that the criterion must not depend on the specific data characteristics.
It must be efficient in order to detect the maximum number of \emph{true}
sources with the minimum error (both, in the
amplitude and position determination). 

Up to this point, we just are able to provide
\emph{cleaned} maps of the sky with good PS amplitude estimation
obtained from a multiscale fit. To decide which one is a
\emph{real} source from the map of maxima is an issue that
is outside the scope of this paper. There are several works in the literature
dealing with
this problem. One of them defines a detection criterion for 
the maxima based on the
acceptation region (using the \emph{Amplitude-Curvature} space)
given by the Neyman-Pearson lemma
(Barreiro et al. 2003); results are presented for backgrounds 
described by Gaussian random fields.
Other authors have suggested the False Discovery
Rate (FDR) method (Hopkins et al. 2002) or a Bayesian approach
to detect and characterize the maxima (Hobson \& McLachlan 2002).
In general the application of
detection criterions having the desirable
properties mentioned above is, in practice, very complicated.
Moreover, they have been applied to situations for which the
background has very simple and well known statistical properties.
Even more, they have been tested on small data sets. 
For the \emph{Planck} case the situation is very different: some of the 
backgrounds have complex statistical properties and the data sets are very 
large.
The results presented in the next Section are obtained
assuming a detection criterion that is able to get a catalogue with
a maximum percentage ($5\%$) of spurious detection, where spurious
is a detection with an error (in absolute value) in the amplitude
estimation larger than $50\%$
(see Vielva et al. 2001a for details). 
This simple detection criterion assumes that the simulations for
the different components describe well enough the main statistical
properties of the \emph{real ones}. Hence, in practice,
this exercise gives us a flux limit which will be used as a
threshold for the \emph{real case}. Above this threshold,
the percentage of spurious detections will be expected to be $\leq 5\%$.
We are working on a detection criterion for the Planck data
that takes into account multi-frequency
information as well as simple PS distribution features.
Results will be presented in a future work.

\section{Results}
\begin{figure}
%
%
	\begin{center}

%
%
		\includegraphics[angle=270,width=8cm]{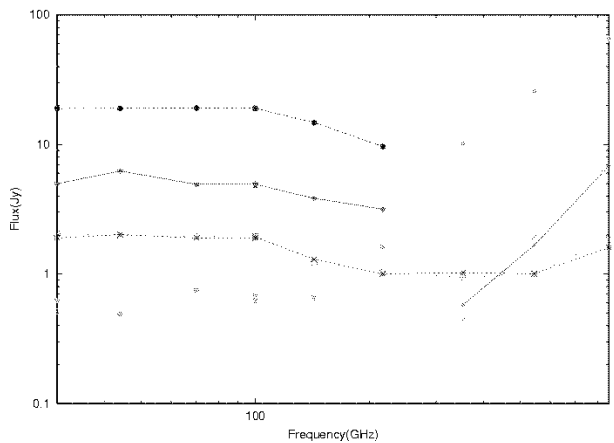}

\caption{\label{indice}
Spectral behaviour for some of the detected sources found in
coincidence at several channels.}
\end{center}
\end{figure}

\begin{table*}
%
%
   \begin{center}
         \begin{tabular}{|c|c|c|c|c|c|c|c|}
	 \hline
	 Frequency (GHz)& \# & Min Flux (Jy) & 
         $\bar{E}(\%)$  & $\bar{b}(\%)$ & Galactic Cut(deg) &
         $N_{R_o}$ & Completeness (\%)\\
	 \hline
	 857      & 27257 & 0.48 & 17.7 &  -4.4 & 25   & 17 & 70 \\
	 545      &  5201 & 0.49 & 18.7 &   4.0 & 15   & 15 & 75 \\
	 353      &  4195 & 0.18 & 17.7 &   1.4 & 10   & 10 & 70 \\
	 217      &  2935 & 0.12 & 17.0 &  -2.5 &  7.5 &  4 & 80 \\
	 143      &  3444 & 0.13 & 17.5 &  -4.3 &  2.5 &  2 & 90 \\
	 100(HFI) &  3342 & 0.16 & 16.3 &  -7.0 &  0   &  4 & 85 \\
	 100(LFI) &  2728 & 0.19 & 17.0 &  -2.4 &  0   &  4 & 80 \\
	  70      &  2172 & 0.24 & 17.1 &  -6.7 &  0   &  6 & 80 \\
	  44      &  1987 & 0.25 & 16.4 &  -6.4 &  0   &  9 & 85 \\
	  30      &  2907 & 0.21 & 18.7 &   1.2 &  0   &  7 & 85 \\
   \hline	
   \end{tabular}
   \caption{\label{catalogue}
		PS catalogue obtained from the Planck data using
            the SMHW method proposed.
            In column 2 we print the number of detections 
	    above the flux limit given in column 3 with a maximum 
            $5\%$ of spurious detections.
	    The mean error (in absolute value) is shown in
            column 4, whereas the mean bias is in column 5. The
            Galactic cut below which no detection is available is
            shown in the sixth column. The number of optimal
            scales appears in the column number 7 and the completeness
            of the catalogue in the last one.}
   \end{center}
\end{table*}
\begin{figure*}
%
%
	\begin{center}
		\includegraphics[width=16cm]{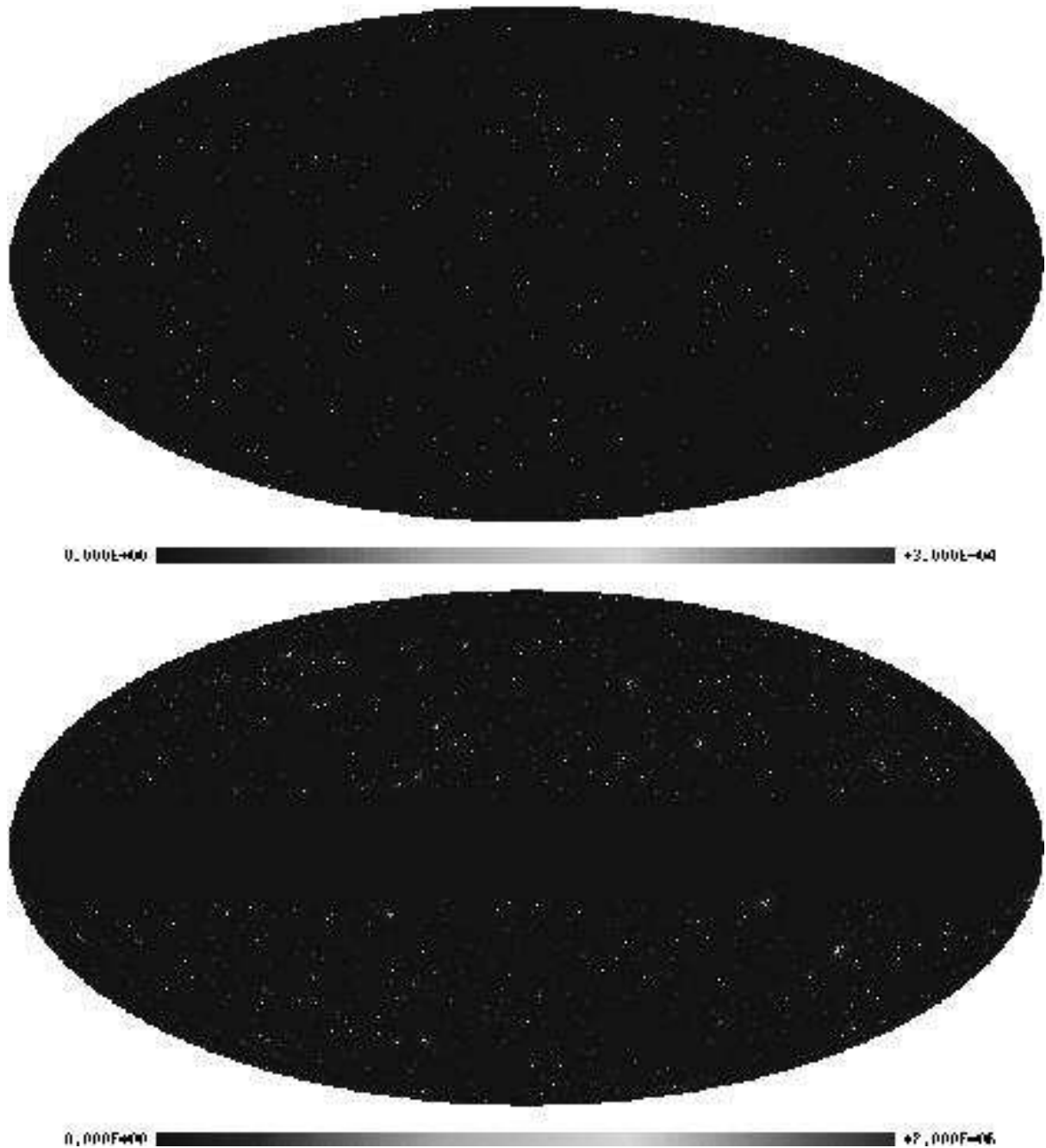}

        \caption{\label{psDetected1}
Recovered point source maps. The frequencies are 44
GHz and 545 GHz. The maps are filtered with a Gaussian beam
(see Table \ref{planck}).}

\end{center}
\end{figure*}
\begin{table*}
%
%
   \begin{center}
         \begin{tabular}{|c|c|c|c|c|c|c|c|c|c|c|c|c|c|c|c|c|}
	 \hline
	      &  
               &    & $85\%$     &           &
               &    & $90\%$     &           &
               &    & $95\%$     &           &
	       &    & $99\%$     &  \\
	 \hline
	 Freq  & 
	 \#  & flux & $\bar{E}$  & $\bar{b}$ & 
	 \#  & flux & $\bar{E}$  & $\bar{b}$ & 
	 \#  & flux & $\bar{E}$  & $\bar{b}$ & 
	 \#  & flux & $\bar{E}$  & $\bar{b}$ \\
	 (GHz) &
	 (Jy)  &    & (\%)       & (\%) &
	 (Jy)  &    & (\%)       & (\%) &
	 (Jy)  &    & (\%)       & (\%) &
	 (Jy)  &    & (\%)       & (\%) \\
	 \hline
	 857      &
	   12763  & 0.78 & 11.1 &   1.1 & 
	   10001  & 0.91 &  9.7 &   1.5 & 
	    6763  & 1.19 &  7.8 &   1.7 & 
	    2816  & 2.00 &  4.7 &   1.1 \\

	 545      &
	    3668  & 0.60 & 16.3 &   5.7 & 
	    2806  & 0.72 & 14.5 &   6.4 & 
	    1844  & 0.91 & 11.7 &   6.3 & 
	     421  & 2.22 &  4.9 &   2.8 \\

	 353      &
	    2233  & 0.28 & 13.4 &   6.2 & 
	    1718  & 0.33 & 12.0 &   6.2 & 
	    1177  & 0.42 & 10.0 &   6.0 & 
	     294  & 0.94 &  4.8 &   3.1 \\

	 217      &
	    2288  & 0.15 & 14.5 &   1.0 & 
	    1850  & 0.18 & 12.9 &   2.6 & 
	    1440  & 0.22 & 11.1 &   3.4 & 
	     779  & 0.34 &  7.8 &   3.8 \\

	 143      &
	   3444n  & 0.13n & 17.5n & -3.3n    & 
	   3444n  & 0.13n & 17.5n & -3.3n    & 
	    2576  & 0.17  & 14.2  & -0.9 & 
	    1776  & 0.23  & 11.4  &  0.2 \\

	 100      &
	    3342n & 0.16n & 16.3n &  -7.0 & 
	    2727  & 0.20  & 13.4  &  -4.5 & 
	    2078  & 0.26  & 10.9  &  -2.6 & 
	    1224  & 0.40  &  7.5  &  -0.7 \\

	 (HFI) \\

	 100      &
	    2370  & 0.22 & 15.2 &  -0.6 & 
	    1768  & 0.29 & 12.7 &   1.6 & 
	    1323  & 0.37 & 10.4 &   3.1 & 
	     635  & 0.65 &  743 &   4.3 \\

	 (LFI) \\

	  70      &
	    2076  & 0.25 & 16.4 &  -5.9 & 
	    1399  & 0.35 & 12.3 &  -2.3 & 
	     867  & 0.51 &  8.2 &   0.2 & 
	     554  & 0.70 &  6.2 &   0.6 \\

	  44      &
	    1987n & 0.25n & 16.4n &  -6.4n& 
	    1489  & 0.33  & 13.2  &  -3.8 & 
	    1092  & 0.43  & 10.6  &  -2.4 & 
	     608  & 0.67  &  7.5  &   1.3 \\

	  30      &
	    2706  & 0.22 & 18.1 &   1.7   & 
	    2264  & 0.26 & 15.4 &   3.7 & 
	    1921  & 0.30 & 14.3 &   3.9 & 
	    1099  & 0.45 & 10.6 &   3.5 \\

   \hline	
   \end{tabular}
   \caption{\label{CatalogueCompleteness}
	PS catalogues for different completeness levels 
	   ($85\%$, $90\%$, $95\%$ and $99\%$). For each 
           completeness level we show the number of detections,
	   the minimum flux and the mean error and bias. The letter
           'n' means that the results on that column are the same than
           in the previous Table.}	
   \end{center}
\end{table*}
\label{results}
Applying the method described in Section \ref{method} to the
simulations shown in Section \ref{simul}, we are
able to obtain all-sky Planck PS catalogues. The catalogue
is presented in Table \ref{catalogue}.
The total number of PS detected is given in the second column. We are
just able to detect those sources in the tail of the PS distribution
(we comment more about this below). 
The minimum fluxes achieved at each frequency are shown in the third
column. At intermediate frequencies we are able to reach lower fluxes 
because it is the \emph{microwave window} where 
Galactic emission is lower. The 100 GHz HFI channel is
specially important for the PS
detection, since it is expected to have a low
instrumental noise. In columns four and five the mean error,
$E(\%)$, and bias, $b(\%)$,
are presented, where:
\begin{eqnarray}
\bar{E}(\%) & = & \sum_{\vec{n}} \frac{|I_{\vec{n}} - \tilde{I}_{\vec{n}}|}{I_{\vec{n}}}100 \\
\bar{b}(\%) & = & \sum_{\vec{n}} \frac{I_{\vec{n}} - \tilde{I}_{\vec{n}}}{I_{\vec{n}}}100,
\end{eqnarray}
being $I_{\vec{n}}$ the PS intensity at location ${\vec{n}}$ and
$\tilde{I}_{\vec{n}}$ the method estimation.
The mean error is between $\approx 16-19\%$, whereas the bias
behaviour changes along the frequency range.
One of the reasons for that change arises from the fact that
we perform a zero-level recalculation of the wavelet coefficients
at high frequency Planck channels.
The new zero-level is not perfectly estimated and, hence,
the PS amplitudes are systematically underestimated or overestimated.
However,
this kind of bias can be taken into account and it can be corrected
in the PS amplitude estimation.
Due to the Galactic emission, there are latitudes where the PS 
detection is
not possible. We show this in column six. At low and intermediate
frequencies the Galaxy is not a problem, but, obviously
at higher frequencies it becomes a handicap. In column seven we
present the number of optimal scales needed in the algorithm. Since
the homogeneous CMB emission is the dominant contribution
at intermediate frequencies, the number of required optimal scales
at these channels is lower than in the other due to the
gradient pattern for the Galactic emission.
In Fig \ref{OptimalScales} we plot, as an example,
the optimal scale maps for the
44 , 100 (LFI), 217 and 545 GHz channels.
Finally, in column eight we represent the completeness
percentage (above the minimum flux and outside the Galactic cut)
that the detected catalogue
represents. The completeness percentage is $\sim 80\%$ for low
and intermediate frequency channels and it decreases for high ones.
In Figs. \ref{psDetected1} (44 and 545 GHz)
we show the PS catalogue
maps,  where the areas without PS detection can be seen
for channels at high frequency
\begin{figure*}
%
%
	\begin{center}

%
%

%
%
		\includegraphics[width=16cm]{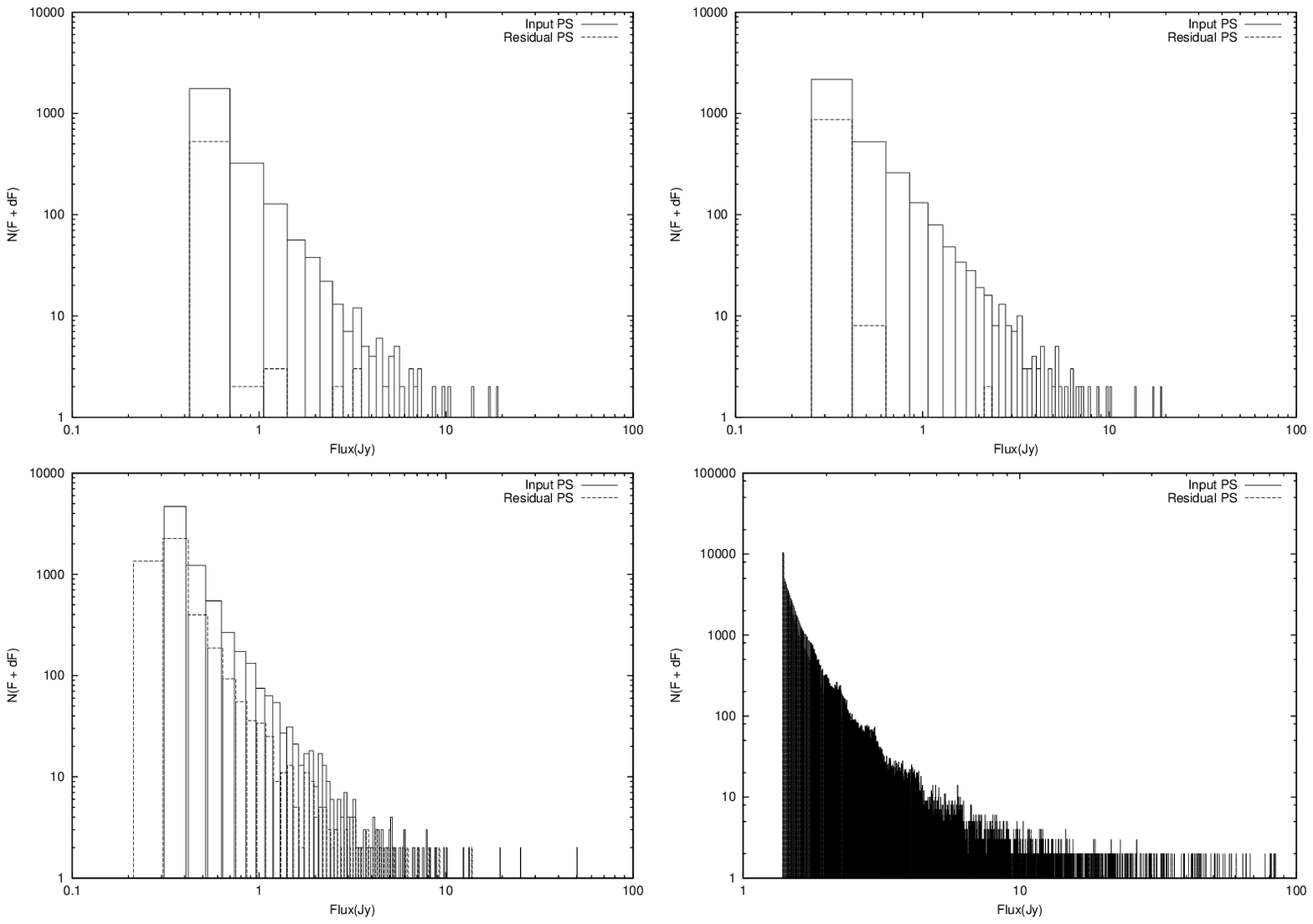}

\caption{\label{histo}
Tails PS distribution function - $N(F)dF$- for the input PS maps
(dark histograms) and the residual ones (light histograms)
obtained by subtracting the
recovered PS catalogue show in Table \ref{catalogue} from the
simulated PS.The frequencies are (from
up-left to bottom-right): 30 GHz, 100 GHz (LFI), 353 GHz and 857 GHz.
}

\end{center}
\end{figure*}
We want to remark that these results have been obtained without
including the rotational dust emission. If this emission is present
in the simulations, the 30 and 44 GHz results change, whereas the
results for the rest of the channels are the same.
In particular, at 30 GHz, the number of 
detected PS is 1809 and the minimum flux is 0.41 Jy; in
addition, 11 optimal scales are required. For the 44 GHz channel
1055 PS are detected and 0.51 Jy is the minimum flux that
we are able to achieve (the number of optimal scales is the same).
An interesting question is to study the different completeness
catalogues that can be achieved with the method.
We present in
Table \ref{CatalogueCompleteness} the complete catalogues at
$85\%$, $90\%$, $95\%$ and $99\%$ levels. For 30 GHz 
and 44 GHz
channels the completeness catalogues at $85\%$ are the same
that the ones in Table \ref{catalogue}, whereas the for
143 GHz channel the catalogue in Table \ref{catalogue}
represents a $90\%$ complete one.
Obviously as the completeness level increases, the number of 
detected PS is lower and the flux limit achieved is higher.
However, the accuracy of the catalogues is better
as the mean errors and bias show.
If we subtract the recovered PS catalogues from the input
PS maps, we obtain \emph{residual} PS maps, with RMS values
lower than the input ones (see Table \ref{rmsps}).
The power spectrum of a Poisson distribution in the sky
of PS is flat and is proportional to the $\textrm{RMS}^2$
\footnote{This is true since we adopted a Poisson spatial distribution
of extra-galactic PS, i.e no clustering of sources has been taken into
account in the present simulations. However, in the case of clustered
sources, a factor very close to the above one is found, at least for
all realistic clustering scenarios at frequencies where flat-spectrum
sources dominate the number counts.}
and therefore the power spectra of the residual PS maps
are correspondingly reduced by a factor  
$(\textrm{RMS}_{\textrm{input}} - \textrm{RMS}_{\textrm{residual}}/
\textrm{RMS}_{\textrm{input}})^2$.
\begin{table}
%
%
   \begin{center}
         \begin{tabular}{|c|c|c|}
	 \hline
	 Frequency & Input PS   & Residual PS  \\
	 (GHz)     &  &   \\ 
	 \hline
	 857       & $1.96\times10^{-2}$ 
	& $1.60\times10^{-2}$ \\
	 \hline
	 545       & $1.91\times10^{-4}$ 
	& $1.53\times10^{-4}$ \\
	 \hline
	 353       & $1.39\times10^{-5}$ 
	& $1.17\times10^{-5}$ \\
	 \hline
	 217       & $3.45\times10^{-6}$ 
	& $2.41\times10^{-6}$ \\
	 \hline
	 143       & $3.54\times10^{-6}$ 
	& $1.65\times10^{-6}$ \\
	 \hline
	 100 (HFI) & $5.41\times10^{-6}$ 
	& $1.30\times10^{-6}$ \\
	 \hline
	 100 (LFI) & $5.79\times10^{-6}$ 
	& $1.54\times10^{-6}$ \\
	 \hline
	 70        & $7.43\times10^{-6}$ 
	& $2.16\times10^{-6}$ \\
	 \hline
	 44        & $1.08\times10^{-5}$ 
	& $3.15\times10^{-6}$ \\
	 \hline
	 30        & $1.54\times10^{-5}$ 
	& $4.56\times10^{-6}$ \\
	 \hline
      \end{tabular}
      \caption{\label{rmsps}The PS RMS values (in $\Delta T/T$ units) for the 
               simulated Planck channels are
               presented. In the second column we show the RMS
               of the input PS maps, whereas in the third one
               it is shown the RMS after the detected PS (the ones in
               Table~\ref{catalogue})
	       are subtracted. All the maps have been convolved with
               the beams.}
    \end{center}
\end{table}
As we commented before, the detected PS belong to the tail of the PS
distributions. This can be seen in Fig. \ref{histo}
where we plot the input and residual
PS tail distribution for the catalogue of Table \ref{catalogue}.
Only the brightest sources
are removed from the map. However, removing these brightest sources
is a critical issue if we want to apply this PS detection method
in combination with
all-component separation ones such as MEM
or Fast Independent Component Analysis (FastICA).
These methods
need to assume that the PS distribution is Gaussian (more realistic
PS distributions require non-Gaussian probability distribution function)
in order
to deal with the PS emission. Clearly, this assumption is closer
to reality if the brightest sources are removed. Moreover, as the
RMS PS decreases, this component is less important (see Vielva
et al. 2001b).
The last result we comment is that the recovered PS catalogues
provide spectral information of the PS populations, since
several PS can be detected at different channels.
Therefore we can study the frequency dependence
of the PS emission, this can be useful to study the physical
processes that generate such emission. This is important
at all Planck channels, because there is a lack of information
at these frequencies. It is particularly relevant 
at intermediate Planck frequencies, where
the knowledge of the different PS populations is really poor.
We can follow 57 sources in the whole Planck spectral interval,
1585 can be seen from 30 GHz to 100 GHz channel, 942 from
143 GHz to 353 GHz and 2231 from 353 GHz to 857 GHz. In
Fig. \ref{indice} we plot some of these sources.
If we focus on spectral intervals small enough to assume
a power law for the PS emission, we can estimate
the spectral indices:
\begin{equation}
F = F_o \Big({\frac{\nu}{\nu_o}}\Big)^{\alpha},
\end{equation}
where $\alpha$ is the spectral index that will define the different PS
populations.
In Table \ref{index} we give an estimation of the spectral indices
for different frequency ranges (and using the catalogue
described in Table \ref{catalogue}). By fitting the PS estimated
amplitudes, we can determine the $\alpha$ parameter that appears
in columns 6, 7 and 8 whereas the input (simulated) values are shown
in columns 3, 4 and 5.
The mean errors are presented in the last column.
For spectral indexes close to zero -- i.e. indexes of sources
associated to QSOs, blazars and AGNs, whose emission is dominated by
the nuclear activity -- we give the absolute error whereas
relative errors are provided for typical spectral indexes of sources
whose emission is dominated by cold dust at microwave frequencies 
(e.g., ultraluminous IRAS galaxies, high redshift spheroids and
starburst/interacting galaxies). From Table \ref{index} it is apparent that
we are able to well recover the average input spectral index in every
frequency range of the Planck experiment.
%

%
A handful of sources (57) are detected in all channels. Albeit this
is a small number if compared to the ones in the complete catalogues,
it will allow to follow the spectral behaviour in a range in which
should appear the steepening of the synchrotron spectrum due to electron
energy losses. This will help to establish, e.g., the synchrotron age of
the source, the relative importance of the free--free emission, etc.
Moreover, the high number
of detected sources in at least two/three nearby channels allows
us to single out the different source populations dominating the 
counts of bright sources at different frequencies. 
On one side, by  the HFI channels, it will be possible to define much
better the spectral properties of the cold dust emission. This information 
-- complemented by the one coming from the surveys of the Herschel 
satellite at higher frequencies -- shall help to fix the temperature of
the cold and warm dust components in different sources. Therefore,
it should be possible, in principle, to distinguish
between high redshift spheroids in the early phase of
their star formation and low redshift normal spirals whose emission is
mostly dominated by the cold dust spread all over the galaxy.
On the other side, a complete catalogue of approximately a thousand
of sources -- detected by LFI and mostly associated to AGN -- can be
released and this will
greatly help to better understand the physical processes originating
in the nuclear region of a galaxy and the properties of the emitted
radiation.
Eventually, a number of inverted spectrum sources should show up
at the intermediate Planck channels and this will help to study
the relevance in number counts and the spectral properties of such
source populations as Gigahertz Peaked (GPS) and Advection Dominated 
(ADS) sources which currently are very poorly known (see Toffolatti
et al 1999).
\begin{table*}
    \begin{center}

	\begin{tabular}{|c|c|c|c|c|c|c|c|c|}
	\hline
	Channels (GHz) & \# &
	$\alpha_{min}$ & $\overline{\alpha}$ & $\alpha_{max}$ &
	${\alpha_{min}}^{est}$ & ${\overline{\alpha}}^{est}$ &
	${\alpha_{max}}^{est}$ & Mean Error\\
	\hline
	857 -- 545 & 4132 & -0.11 &   2.43 &   3.49 & -0.66 &   2.41   & 4.41
	 & 12.29$\%$ \\

	857 -- 545 -- 353 & 2231 & -0.46 & 2.16 & 3.50 & -0.52 & 2.58
	& 3.79 & 7.13$\%$ \\

	353 -- 217 -- 143 & 942 & -0.91 & -0.03 & 3.08 & -2.04 & -0.08
	& 3.40 & 0.17 \\

	100 -- 70 -- 44 -- 30 & 1585 & -0.72 & -0.14 & 0.41 & -1.01 &
	-0.17 & 0.63 & 0.11 \\

	44 -- 30 & 1767 & -1.46 & -0.19 & 0.74 & -2.44 & -0.15 & 2.17
	& 0.31 \\
	\hline

\end{tabular}
\caption{\label{index}
Spectral indices for PS found in coincidence at the channels
indicate in column 1. The number of the PS found is given in the
second column. Columns 3, 4 and 5 give the minimum, mean and maximum
values of the spectral indices as calculated from the input PS. The
spectral index estimation is presented in columns 6, 7 and 8. The mean
error in the estimation (absolute error for spectral index close to
zero are given) is shown in the last column.}
\end{center}
\end{table*}
\section{The case of a non-ideal instrument}
\label{beams}

We have also tested the influence of realistic asymmetric beams not only
in the number of detected point sources, but also in the flux limits
achieved and the mean error in the amplitude estimation. 
This a very important point, since the expected beam shapes for Planck
satellite have a asymmetric profiles.
A detailed study about
the influence of this secondary effect, non-uniform
noise distribution and other systematic effects for the
detection of point sources, will be presented
in a future work. Here we just present the basic procedure to adapt
the SMHW method presented in Section~3 to the case where the beam
is asymmetric and non-uniform and $1/f$ noises are considered.
Although the
SMHW is an isotropic wavelet, it can be also adequate to perform the
detection of point sources that show a slight Gaussian
asymmetry . This is precisely the situation for the Planck
beams, where, due to the
scanning strategy, at a given point in the sky, the \emph{effective
beam} is an average of the pure asymmetric antenna over all the
possible orientations. This makes the resultant beam more symmetric
than the original one. Moreover, we have shown that the SMHW itself
is a very promising tool to characterize the \emph{effective
dispersion} ($\sigma_{eff}$) of the realistic
beam, by performing a multi-scale analysis of point-like objects in the
sky that have been convolved with realistic beams following the expected
Planck scanning strategy, (available for the Planck community at
the web site \emph{http://www.mpa-garching.mpg.de/$\sim$planck/}).
Thanks to this multi-scale analysis, we can determine the
$\sigma_{eff}$ at any direction on the sky. 
Let us remark here the main results of this study
for the case of the largest Planck beam asymmetry: the 30 GHz channel.
We have used simulations where realistic beam, 
\footnote{In the previous web site, the used antenna is refered as the
``old beam'' (J. Tauber 2002).}
non-uniform and $1/f$ ($f_{knee} = 100$ mHz) noises are included.
The asymmetric beam is close
to a Gaussian, with an ellipticity of $\sim 20\%$ with a minor
dispersion of
15.47$'$. We found that, on average, the $\sigma_{eff}$ (after the
scanning strategy) is around 1.39 times the nominal Planck dispersion in
this channel.
By comparing the point source catalogue that the SMHW
provides in this realistic case, with the one that we would get
in the case where an ideal Gaussian beam is used (with an area
equal to the one of the real asymmetric beam, that is 1.2 times the
nominal dispersion),
we show that around 80\% of
the point sources are still recovered, and the flux limit is just 10\% above
the one found for the ideal case. The mean error and bias are
practically the same in both cases.
We want to remark that the discrepancy between the effective beam
value average to all the sky, 1.39,
that we found for the realistic antenna and the 1.2 value for the
ideal case (both in units of the
nominal dispersion) arises from the scanning strategy that widens the
effective beam in the realistic situation.
\section{Conclusions and discussion}
\label{fin}
We have extended the work done in Vielva et al. (2001a)
focused on the detection of PS in simulated microwave flat patches
of the sky, to the whole celestial sphere.
We have done a spherical analysis using the
Spherical Mexican Hat Wavelet that is obtained as
an stereographic projection of the plane Mexican Hat Wavelet.
This extension implies several new steps in the MHW detection
method. First of all, we need to estimate the optimal scale
at each area of the sky, in order to achieve the maximum
number of detections. To perform the optimal
scale determination of the spherical maps is a huge task
that involves several transformations (from harmonic domain to wavelet
space and viceversa) 
requires a long CPU time and/or storage space.
Using the fact that the HEALPix package allows to
identify quickly pixels in the sky, we have done
plain projections of HEALPix pixels (at $N_{\rm side}$ = 4)
to calculate the optimal
scale using Fourier Transform. This approach reduces enormously
the CPU time to estimate the optimal scale. We have checked that
the error introduced in the optimal scale determination
due to the projection is lower than $5\%$.
We have tested that the amplification (flux limit)
is greater (lower) than $1\%$ the one obtained using
the exact value of the optimal scale. 
A big effort has
been also done to reduce the CPU time needed for the harmonic
transforms. Whereas the SMHW convolution is done in the
harmonic domain, the PS detection is done in wavelet space.
In addition, three adjacent scales are required for each optimal
one, in order to perform a multiscale fit of the wavelet coefficients
to better estimate the PS amplitudes. We have solved
this problem by modifying some of the HEALPix package codes to perform
several harmonic transformations at the same time using 
pre-computed Legendre polynomials.
Applying the presented algorithm with the
detection criterion based on the simulations
(see Subsection~\ref{detect})
that is able to produce PS catalogues
with a maximum percentage of spurious detections ($5\%$)
--where spurious means an error larger than $50\%$--,
we are able to
recover a PS catalogue for each of the 10 Planck channels
(Table \ref{catalogue}). We can also provide information
of complete PS catalogues at different levels 
($85\%$, $90\%$, $95\%$ and $99\%$). These catalogues are in
Table \ref{CatalogueCompleteness}. By subtracting the
detected PS from the input PS maps, we can construct PS residual
maps that show us how the removed sources are in the high flux
tails of the PS distributions: we are just able to detect the brightest
sources. However, removing these sources is very important
in order to apply all-component separation methods, as MEM
or FastICA. These methods assume that the different component
emissions can be factorized in a spatial template and a 
frequency dependent pattern. This is clearly false for the PS
emission. To deal with that emission, the all-component separation
methods needs to assume the PS signal as a \emph{noisy contribution},
which for simplicity, is assumed to be Gaussian (different
distribution functions for the PS require to add more
assumptions to the method). Although the presented residual PS maps
are not Gaussian, they are closer to Gaussian than
the input PS distribution. In addition, they have RMS values
significantly lower than the RMS input maps, what makes the PS emission less
important (see Vielva et al. 2001b).
The largest number of detections are at high and intermediate
frequencies. At high frequencies the dominant PS
populations are due to galaxies whose emission is dominated by dust whereas,
at low frequencies, almost all the bright detected sources are
flat-spectrum AGN and QSOs. The channels in which the detection
result easier are the intermediate ones, in which the Galactic
emission is at a minimum. In fact, in the maps where the Galactic
emission is relatively more important (at high and low frequencies)
the number of local optimal scales needed increases (see Table 3).
By following the detected PS emission in several channels,
we are able to provide information about the spectral behaviour
of the different PS populations. This is an important result,
since, at present, there is not much information about
the PS emission at the Planck frequencies (specially at intermediate
ones). We are able to follow such emission in several
channels, and a few of them along all the Planck
frequency range. By fitting the estimated PS amplitudes
in small frequency ranges, we can determine sources spectral indices.
That is important in order to study the physical
processes that produce such emission. We can recover the spectral
index with good accuracy as shown in Table \ref{index}.
We would like to remark that the method is robust in the sense that
previous knowledge about the underlaying signals
present in the map are not needed: all the information
to perform the
analysis (the optimal scales) come from the data itself. The only
relevant assumption made in the method is that the beam has a Gaussian
pattern.
An additional support for the robustness of the method is related
to the good detection performance for asymmetric beams,
like the ones expected from
Planck. The influence of these realistic beams
has been tested using the antenna of the Planck
satellite with the largest asymmetry (the LFI 28 beam at 30 GHz).
Even more, the SMHW can be useful to characterise the beam asymmetry,
by permorfing a multi-scale analysis. An specific study concerning
the influence on asymmetric beams, non-uniform
noise distribution and other systematic effects for the
detection of point sources, will be presented
in a future work.
Finally, we plan to combine this method with the MEM and FastICA
spherical algorithms, in order to perform the all-component
separation of the microwave sky.
\section*{Acknowledgments}
We thank the referee of the paper for helpful suggestions and
comments.
We also thank R. B. Barreiro and D. Herranz for useful comments.
PV acknowledges support from Universidad de Cantabria fellowship.
We acknowledge partial financial support from the Spanish MCYT
projects ESP2001-4542-PE and ESP2002-04141-C03-01.
We thank Centro de Supercomputacion de Galicia (CESGA) for providing
the Compaq HPC320 supercomputer and R. Marco for kindly providing
the IFCA computer net GRID to run part of the code.
This work has used the software package HEALPix (Hierarchical, Equal
Area and iso-latitude pixelization of the sphere,
http://www.eso.org/science/healpix), developed by K.M. Gorski,
E. F. Hivon, B. D. Wandelt, J. Banday, F. K. Hansen and M. Barthelmann.

\bsp

\end{document}